%% file: main.tex
\documentclass[sigconf]{acmart}
 
\usepackage{booktabs} 
\usepackage{multirow}
\usepackage{mathtools}
\usepackage{amsmath}
\usepackage{subcaption}
\usepackage{tabularx,color}
\usepackage{enumitem}
\usepackage{appendix}
\usepackage[flushleft]{threeparttable}
\DeclareSymbolFont{extraup}{U}{zavm}{m}{n}
\DeclareMathSymbol{\varheart}{\mathalpha}{extraup}{86}
\DeclareMathSymbol{\vardiamond}{\mathalpha}{extraup}{87}
\newcommand{\xhdr}[1]{{\noindent\bfseries #1}.}

\copyrightyear{2018}
\acmYear{2018}
\setcopyright{acmlicensed}
\acmConference[Woodstock '18]{Woodstock '18: ACM Symposium on Neural Gaze Detection}{June 03--05, 2018}{Woodstock, NY}
\acmPrice{15.00}
\acmDOI{10.1145/1122445.1122456}
\acmISBN{978-1-4503-9999-9/18/06}

\setcopyright{acmcopyright}

\begin{document}
\title{AutoInt: Automatic Feature Interaction Learning via Self-Attentive Neural Networks}

\author{Weiping Song}\authornote{Part of this work was performed when the first author was visiting Mila.}
\affiliation{\institution{Department of Computer Science, School of EECS, Peking University}}
\email{weiping.song@pku.edu.cn}

\author{Chence Shi}
\affiliation{\institution{Department of Computer Science, School of EECS, Peking University}}
\email{chenceshi@pku.edu.cn}

\author{Zhiping Xiao}
\affiliation{\institution{Department of Computer Science, University of California, Los Angeles}}
\email{patriciaxiao@g.ucla.edu}

\author{Zhijian Duan, Yewen Xu}
\affiliation{\institution{Department of Computer Science, School of EECS, Peking University}}
\email{{zjduan,xuyewen}@pku.edu.cn}

\author{Ming Zhang}\authornote{Corresponding authors.}
\affiliation{\institution{Department of Computer Science, School of EECS, Peking University}}
\email{mzhang_cs@pku.edu.cn}

\author{Jian Tang}\authornotemark[2]
\affiliation{\institution{Mila-Quebec AI Institute, \\HEC Montreal \& CIFAR AI Chair}}
\email{jian.tang@hec.ca}




\renewcommand{\shortauthors}{W. Song et al.}

\fancyhead{}

\begin{abstract}
Click-through rate (CTR) prediction, which aims to predict the probability of a user clicking on an ad or an item, is critical to many online applications such as online advertising and recommender systems. The problem is very challenging since (1) the input features (e.g., the user id, user age, item id, item category) are usually sparse and high-dimensional, and (2) an effective prediction relies on high-order combinatorial features (\textit{a.k.a.} cross features), which are very time-consuming to hand-craft by domain experts and are impossible to be enumerated. Therefore, there have been efforts in finding low-dimensional representations of the sparse and high-dimensional raw features and their meaningful combinations.

In this paper, we propose an effective and efficient method called the \emph{AutoInt} to automatically learn the high-order feature interactions of input features. Our proposed algorithm is very general, which can be applied to both numerical and categorical input features. Specifically, we map both the numerical and categorical features into the same low-dimensional space. Afterwards, a multi-head self-attentive neural network with residual connections is proposed to explicitly model the feature interactions in the low-dimensional space. With different layers of the multi-head self-attentive neural networks, different orders of feature combinations of input features can be modeled. 
The whole model can be efficiently fit on large-scale raw data in an end-to-end fashion. Experimental results on four real-world datasets show that our proposed approach not only outperforms existing state-of-the-art approaches for prediction but also offers good explainability.
Code is available at: \url{https://github.com/DeepGraphLearning/RecommenderSystems}.
\end{abstract}

%
%



\copyrightyear{2019} 
\acmYear{2019} 
\acmConference[CIKM '19]{The 28th ACM International Conference on Information and Knowledge Management}{November 3--7, 2019}{Beijing, China}
\acmBooktitle{The 28th ACM International Conference on Information and Knowledge Management (CIKM '19), November 3--7, 2019, Beijing, China}
\acmPrice{15.00}
\acmDOI{10.1145/3357384.3357925}
\acmISBN{978-1-4503-6976-3/19/11}

\maketitle

\input{intro}
\input{related}
\input{definition}

\input{model}
\input{experiment}
\input{conclusion}
\input{ack}

\bibliographystyle{ACM-Reference-Format}
\bibliography{reference}

\end{document}

%% file: intro.tex
\section{Introduction}


Predicting the probabilities of users clicking on ads or items (a.k.a., click-through rate prediction) is a critical problem for many applications such as online advertising and recommender systems~\cite{graepel2010web,he2014practical,cheng2016wide}.
The performance of the prediction has a direct impact on the final revenue of the business providers. Due to its importance, it has attracted growing interest in both academia and industry communities. 

Machine learning has been playing a key role in click-through rate prediction, 
which is usually formulated as supervised learning with user profiles and item attributes as input features. The problem is very challenging for several reasons. First, the input features are extremely sparse and high-dimensional~\cite{mcmahan2013ad,shan2016deep,he2017neural,cheng2016wide,guo2017deepfm}. In real-world applications, a considerable percentage of user's demographics and item's attributes are usually discrete and/or categorical. To make supervised learning methods applicable, these features are first converted to a one-hot encoding vector, which can easily result in features with millions of dimensions. Taking the well-known CTR prediction data Criteo\footnote{http://labs.criteo.com/2014/09/kaggle-contest-dataset-now-available-academic-use/} as an example, the feature dimension is approximately 30 million with sparsity over 99.99\%. With such sparse and high-dimensional input features, the machine learning models are easily overfitted.
Second, as shown in extensive literature~\cite{cheng2016wide,guo2017deepfm,lian2018xdeepfm,shan2016deep}, high-order feature interactions\footnote{In this paper, we will use ``combinatorial feature'' and ``feature interaction'' interchangeably as they are both used in the literature~\cite{shan2016deep,lian2018xdeepfm,guo2017deepfm} .} are crucial for a good performance. For example, it is reasonable to recommend \textit{Mario Bros.}, a famous video game, to \textit{David}, who is a ten-year-old boy. In this case, the third-order combinatorial feature <\textit{Gender=Male, Age=10, ProductCategory=VideoGame}> is very informative for prediction. However, finding such meaningful high-order combinatorial features heavily relies on domain experts. Moreover, it is almost impossible to hand-craft all the meaningful combinations~\cite{rendle2010factorization,cheng2016wide}. One may ask that we can enumerate all the possible high-order features and let machine learning models select the meaningful ones. However, enumerating all the possible high-order features will exponentially increase the dimension and sparsity of the input features, leading to a more serious problem of model overfitting. Therefore, there has been extensive efforts in the communities in finding low-dimensional representations of the sparse and high-dimensional input features and meanwhile modeling different orders of feature combinations. 
For example, Factorization Machines (FM)~\cite{rendle2010factorization}, which combine polynomial regression models with factorization
techniques, are developed to model feature interactions and have been proved effective for various tasks~\cite{rendle2011fast,rendle2010factorizing}. However, limited by its polynomial fitting time, it is only effective for modeling low-order feature interactions and impractical to capture high-order feature interactions. Recently, many works~\cite{he2017neural,cheng2016wide,guo2017deepfm,wang2017deep} based on deep neural networks have been proposed to model the high-order feature interactions. Specifically, multiple layers of non-linear neural networks are usually used to capture the high-order feature interactions. However, such kinds of methods suffer from two limitations. First, fully-connected neural networks have been shown inefficient in learning multiplicative feature interactions~\cite{beutel2018latent}.
Second, since these models learn the feature interactions in an implicit way, they lack good explanation on which feature combinations are meaningful. Therefore, we are looking for an approach that is able to explicitly model different orders of feature combinations, represent the entire features into low-dimensional spaces, and meanwhile offer good model explainability. 



In this paper, we propose such an approach based on the multi-head self-attention mechanism~\cite{vaswani2017attention}. Our proposed approach learns effective low-dimensional representations of the sparse and high-dimensional input features and is applicable to both the categorical and/or numerical input features. 
Specifically, both the categorical and numerical features are first embedded into low-dimensional spaces, which reduces the dimension of the input features and meanwhile allows different types of features to interact with each other via vector arithmetic (e.g., summation and inner product). 
Afterwards, we propose a novel interacting layer to promote the interactions between different features. 
Within each interacting layer, each feature is allowed to interact with all the other features and is able to automatically identify relevant features to form meaningful higher-order features via the multi-head attention mechanism~\cite{vaswani2017attention}. Moreover, the multi-head mechanism projects a feature into multiple subspaces, and hence it can capture different feature interactions in different subspaces. Such an interacting layer models the one-step interaction between the features. By stacking multiple interacting layers, we are able to model different orders of feature interactions. In practice, the residual connection~\cite{he2016deep} is added to the interacting layer, which allows combining different orders of feature combinations. We use the attention mechanism for measuring the correlations between features, which offers good model explainability.






To summarize, in this paper we make the following contributions:
\begin{itemize}
\item We propose to study the problem of explicitly learning high-order feature interactions and meanwhile finding models with good explainability for the problem. 
\item We propose a novel approach based on self-attentive neural network, which can automatically learn high-order feature interactions and efficiently handle large-scale high-dimensional sparse data. 
\item We conducted extensive experiments on several real-world data sets. Experimental results on the task of CTR prediction show that our proposed approach not only outperforms existing state-of-the-art approaches for prediction but also offers good model explainability. 
\end{itemize}

Our work is organized as follows. In Section 2, we summarize the related work. Section 3 formally defines our problem. Section 4 presents the proposed approach to learn feature interactions. In Section 5, we present the experimental results and detailed analysis. We conclude this paper and point out the future work in Section 6.

%% file: related.tex
\section{Related work}
Our work is relevant to three lines of work: 1) Click-through rate prediction in recommender systems and online advertising, 2) techniques for learning feature interactions, and 3) self-attention mechanism and residual networks in the literature of deep learning. 

\subsection{Click-through Rate Prediction}


Predicting click-through rates is important to many Internet companies, and various systems have been developed by different companies~\cite{richardson2007predicting,graepel2010web,mcmahan2013ad,he2014practical,cheng2016wide,covington2016deep,zhou2017deep}. For example, Google developed the Wide\&Deep\cite{cheng2016wide} learning system for recommender systems, which combines the advantages of both the linear shallow models and deep models. The system achieves remarkable performance in APP recommendation. The problem also receives a lot of attention in the academic communities. For example, \citet{shan2016predicting} proposed a context-aware CTR prediction method which factorized three-way <user, ad, context> tensor. \citet{oentaryo2014predicting} developed hierarchical importance-aware factorization machine to model dynamic impacts of ads. 

\subsection{Learning Feature Interactions}
Learning feature interactions is a fundamental problem and therefore extensively studied in the literature. A well-known example is Factorization Machines (FM)~\cite{rendle2010factorization}, which were proposed to mainly capture the first- and second-order feature interactions and have been proved effective for many tasks in recommender systems~\cite{rendle2010factorizing,rendle2011fast}. Afterwards, different variants of factorization machines have been proposed. For example, Field-aware Factorization Machines (FFM)~\cite{juan2016field}  modeled fine-grained interactions between features from different fields. GBFM~\cite{cheng2014gradient} and AFM~\cite{xiao2017attentional} considered the importance of different second-order feature interactions. However, all these approaches focus on modeling low-order feature interactions. 

There are some recent works that model high-order feature interactions. For example, NFM~\cite{he2017neural} stacked deep neural networks on top of the output of the second-order feature interactions to model higher-order features. Similarly, PNN~\cite{qu2016product}, FNN~\cite{zhang2016deep}, DeepCrossing~\cite{shan2016deep}, Wide\&Deep~\cite{cheng2016wide} and DeepFM~\cite{guo2017deepfm} utilized feed-forward neural networks to model high-order feature interactions. However, all these approaches learn the high-order feature interactions in an implicit way and therefore lack good model explainability.  
On the contrary, there are three lines of works that learn feature interactions in an explicit fashion. First, Deep\&Cross~\cite{wang2017deep} and xDeepFM~\cite{lian2018xdeepfm} took outer product of features at the bit- and vector-wise level respectively. Although they perform explicit feature interactions, it is not trivial to explain which combinations are useful. 
Second, some tree-based methods~\cite{zhu2017deep,zhao2017gb,wang2018tem} combined the power of embedding-based models and tree-based models but had to break training procedure into multiple stages. Third, HOFM~\cite{blondel2016higher} proposed efficient training algorithms for high-order factorization machines. However, HOFM requires too many parameters and only its low-order (usually less than 5) form can be practically used.
Different from existing work, we explicitly model feature interactions with attention mechanism in an end-to-end manner, and probe the learned feature combinations via visualization.


\subsection{Attention and Residual Networks} 
Our proposed model makes use of the latest techniques in the literature of deep learning: attention~\cite{bahdanau2014neural} and residual networks~\cite{he2016deep}. Attention is first proposed in the context of neural machine translation~\cite{bahdanau2014neural} 
and has been proved effective in a variety of tasks such as question answering~\cite{sukhbaatar2015end}, text summarization~\cite{rush2015neural}, and recommender systems~\cite{zhou2017deep, song2019session,he2018nais}. \citet{vaswani2017attention} further proposed multi-head self-attention to model complicated dependencies between words in machine translation.

Residual networks~\cite{he2016deep} achieved state-of-the-art performance in the ImageNet contest. Since the residual connection, which can be simply formalized as $y=F(x)+x$, encourages gradient flow through interval layers, it becomes a popular network structure for training very deep neural networks.

%% file: definition.tex
\section{Problem Definition}
We first formally define the problem of click-through rate (CTR) prediction as follows:

\textup{DEFINITION 1.} (\textbf{CTR Prediction}) Let $\mathbf{x}\in \mathbb{R}^n$ denotes the concatenation of user $u$'s features and item $v$'s features, where categorical features are represented with one-hot encoding, and $n$ is the dimension of concatenated features. The problem of \textit{click-through rate prediction} aims to predict the probability of user $u$ clicking on item $v$ according to the feature vector $\mathbf{x}$.

A straightforward solution for CTR prediction is to treat $\mathbf{x}$ as the input features and deploy the off-the-shelf classifiers such as logistic regression. However, since the original feature vector $\mathbf{x}$ is very sparse and high-dimensional, the model will be easily overfitted. Therefore, it is desirable to represent the raw input features in low-dimensional continuous spaces. Moreover, as shown in existing literature, it is crucial to utilize the higher-order combinatorial features to yield good prediction performance~\cite{rendle2010factorization,cheng2016wide,shan2016deep,novikov2016exponential,guo2017deepfm,blondel2016polynomial}. Specifically, we define the high-order combinatorial features as follows:

\textup{DEFINITION 2.} (\textbf{p-order Combinatorial Feature}) Given input feature vector $\mathbf{x}\in \mathbb{R}^n$, a \textit{p-order combinatorial feature} is defined as 
$g(x_{i_1}, ..., x_{i_p})$
, where each feature comes from a distinct field, $p$ is the number of involved feature fields, and $g(\cdot)$ is a non-additive combination function, such as multiplication~\cite{rendle2010factorization} and outer product~\cite{lian2018xdeepfm,wang2017deep}. For example, $x_{i_1}\times x_{i_2}$ is a second-order combinatorial feature involving $x_{i_1}$ and $x_{i_2}$.

Traditionally, meaningful high-order combinatorial features are hand-crafted by domain experts. However, this is very time-consuming and hard to generalize to other domains. Besides, it is almost impossible to hand-craft all meaningful high-order features. Therefore, we aim to develop an approach that is able to automatically discover the meaningful high-order combinatorial features and meanwhile map all these features into low-dimensional continuous spaces. Formally, we define our problem as follows:


\textup{DEFINITION 3.} (\textbf{Problem Definition}) Given an input feature vector $\mathbf{x}\in \mathbb{R}^n$ for click-through rate prediction, our goal is to learn a low-dimensional representation of $\mathbf{x}$, which models the high-order combinatorial features.

%% file: model.tex
\section{AutoInt: Automatic Feature Interaction Learning}\label{sec::model}

In this section, we first give an overview of the proposed approach \textit{AutoInt}, which can automatically learn feature interactions for CTR prediction. Next, we present a comprehensive description of how to learn a low-dimensional representation that models high-order combinatorial features without manual feature engineering.

\begin{figure}
\includegraphics[width=\linewidth]{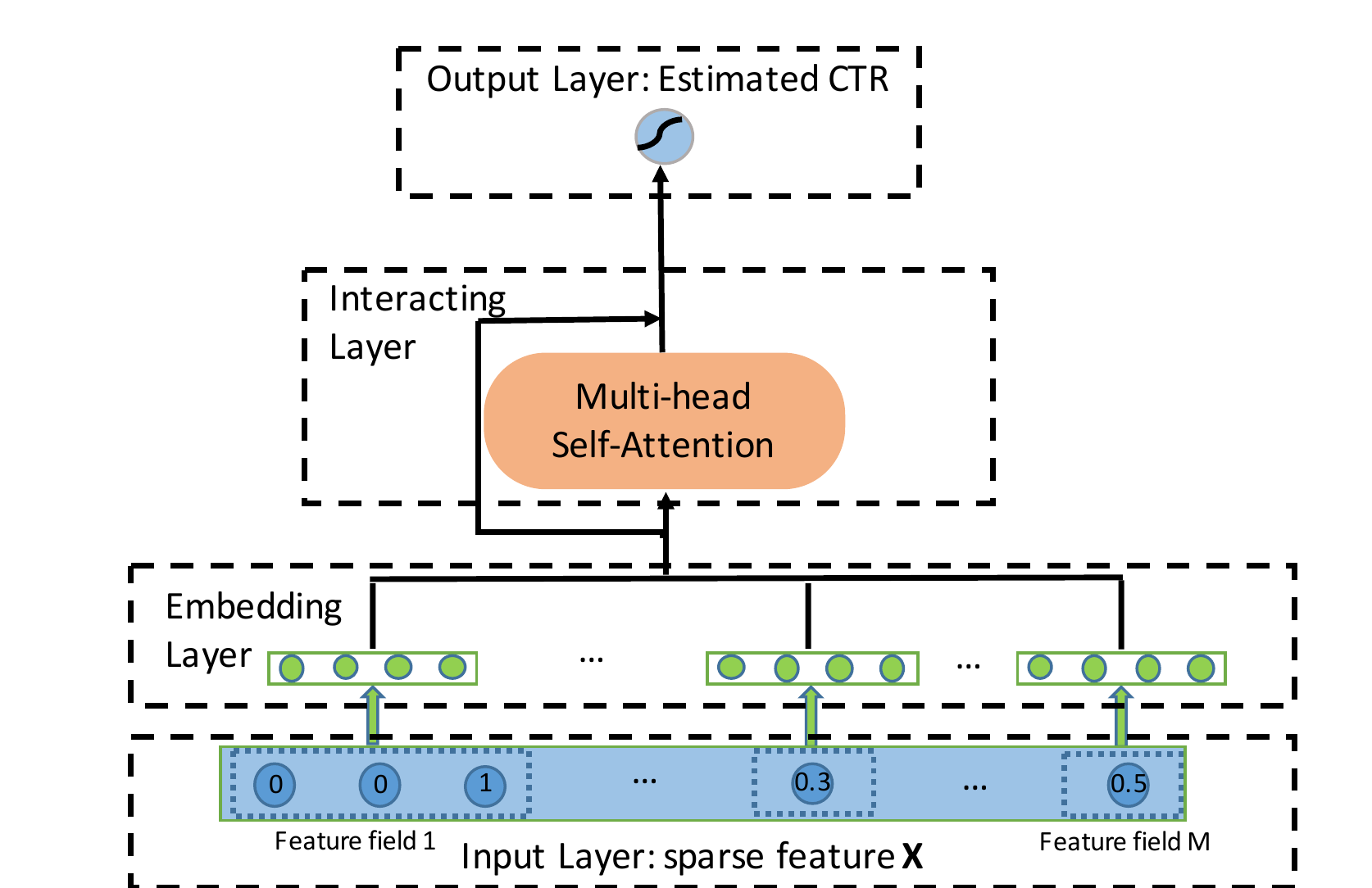}
\caption{Overview of our proposed model AutoInt. The details of embedding layer and interacting layer are illustrated in Figure~\ref{fig::embedding_layer} and Figure~\ref{fig::attention} respectively.}
\label{fig::overview}
\end{figure}

\subsection{Overview}
The goal of our approach is to map the original sparse and high-dimensional feature vector into low-dimensional spaces and meanwhile model the high-order feature interactions. 
As shown in Figure~\ref{fig::overview}, our proposed method takes the sparse feature vector $\mathbf{x}$ as input, followed by an embedding layer that projects all features (i.e., both categorical and numerical features) into the same low-dimensional space. Next, we feed embeddings of all fields into a novel interacting layer, which is implemented as a multi-head self-attentive neural network. For each interacting layer, high-order features are combined through the attention mechanism, and different kinds of combinations can be evaluated with the multi-head mechanisms, which map the features into different subspaces. By stacking multiple interacting layers, different orders of combinatorial features can be modeled.

The output of the final interacting layer is the low-dimensional representation of the input feature, which models the high-order combinatorial features and is further used for estimating the click-through rate through a sigmoid function. Next, we introduce the details of our proposed method.


\subsection{Input Layer}\label{sec::input_layer}

\begin{figure}
\includegraphics[width=\linewidth]{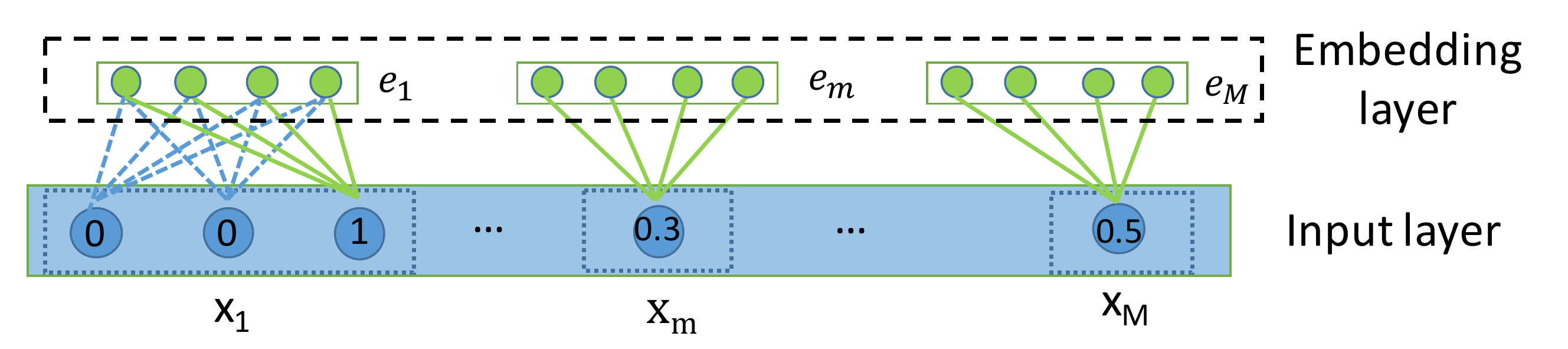}
\caption{Illustration of input and embedding layer, where both categorical and numerical fields are represented by low-dimensional dense vectors.}\label{fig::embedding_layer}\vspace{-10pt}
\end{figure}

We first represent user's profiles and item's attributes as a sparse vector, which is the concatenation of all fields. Specifically,
\begin{equation}\label{eqa::input}
\mathbf{x} = [\mathbf{x_1}; \mathbf{x_2}; ...; \mathbf{x_M}],
\end{equation}
where $M$ is the number of total feature fields, and $\mathbf{x_i}$ is the feature representation of the $i$-th field. $\mathbf{x_i}$ is a one-hot vector if the $i$-th field is categorical (e.g., $\mathbf{x_1}$ in Figure~\ref{fig::embedding_layer}). $\mathbf{x_i}$ is a scalar value if the $i$-th field is numerical (e.g., $\mathbf{x_M}$ in Figure~\ref{fig::embedding_layer}).

\subsection{Embedding Layer}
Since the feature representations of the categorical features are very sparse and high-dimensional, a common way is to represent them into low-dimensional spaces (e.g., word embeddings). Specifically, we represent each categorical feature with a low-dimensional vector, i.e.,  
\begin{equation}\label{eq::categorical_feature}
\mathbf{e_i} = \mathbf{V_i}\mathbf{x_i},
\end{equation}
where $\mathbf{V_i}$ is an embedding matrix for field $i$, and $\mathbf{x_i}$ is an one-hot vector. Often times categorical features can be multi-valued, i.e., $\mathbf{x_i}$ is a multi-hot vector. Take movie watching prediction as an example, there could be a feature field \emph{Genre} which describes the types of a movie and it may be multi-valued (e.g., Drama and Romance for movie ``Titanic''). To be compatible with multi-valued inputs, we further modify the Equation~\ref{eq::categorical_feature} and represent the multi-valued feature field as the average of corresponding feature embedding vectors:
\begin{equation}
\mathbf{e_i} = \frac{1}{q}\mathbf{V_i}\mathbf{x_i},
\end{equation}
where $q$ is the number of values that a sample has for $i$-th field and $\mathbf{x_i}$ is the multi-hot vector representation for this field.

To allow the interaction between categorical and numerical features, we also represent the numerical features in the same low-dimensional feature space. Specifically, we represent the numerical feature as
\begin{equation}
\mathbf{e_m} = \mathbf{v_m}x_m,
\end{equation}
where $\mathbf{v_m}$ is an embedding vector for field $m$, and $x_m$ is a scalar value. 

By doing this, the output of the embedding layer would be a concatenation of multiple embedding vectors, as presented in Figure~\ref{fig::embedding_layer}.



\subsection{Interacting Layer}
Once the numerical and categorical features live in the same low-dimensional space, we move to model high-order combinatorial features in the space. The key problem is to determine which features should be combined to form meaningful high-order features. Traditionally, this is accomplished by domain experts who create meaningful combinations based on their knowledge. In this paper, we tackle this problem with a novel method, the multi-head self-attention mechanism~\cite{vaswani2017attention}. 

Multi-head self-attentive network~\cite{vaswani2017attention} has recently achieved remarkable performance in modeling complicated relations. For example, it shows superiority for modeling arbitrary word dependency in machine translation~\cite{vaswani2017attention} and sentence embedding~\cite{lin2017structured}, and has been successfully applied to capturing node similarities in graph embedding~\cite{velickovic2017graph}. Here we extend this latest technique to model the correlations between different feature fields.

Specifically, we adopt the key-value attention mechanism~\cite{miller2016key} to determine which feature combinations are meaningful. Taking the feature $m$ as an example, next we explain how to identify multiple meaningful high-order features involving feature $m$.  We first define the correlation between feature $m$ and feature $k$ under a specific attention head $h$ as follows:
\begin{equation}\label{eqa::align}
\begin{gathered}
\mathbf{\alpha_{m,k}^{(h)}} = \frac{\exp(\psi^{(h)}(\mathbf{e_m}, \mathbf{e_k}))}{\sum_{l=1}^{M}\exp(\psi^{(h)}(\mathbf{e_m}, \mathbf{e_l}))},  \\
\psi^{(h)}(\mathbf{e_m}, \mathbf{e_k}) = \langle \mathbf{W^{(h)}_{Query}}\mathbf{e_m}, \mathbf{W^{(h)}_{Key}}\mathbf{e_k}  \rangle,
\end{gathered}
\end{equation}
where $\psi^{(h)}(\cdot, \cdot)$ is an attention function which defines the similarity between the feature $m$ and $k$. It can be defined as a neural network or as simple as inner product, i.e., $\langle\cdot,\cdot\rangle$. In this work, we use inner product due to its simplicity and effectiveness. $\mathbf{W^{(h)}_{Query}}$, $\mathbf{W^{(h)}_{Key}}\in \mathbb{R}^{d'\times d}$ in Equation~\ref{eqa::align} are transformation matrices which map the original embedding space $\mathbb{R}^{d}$ into a new space $\mathbb{R}^{d'}$. Next, we update the representation of feature $m$ in subspace $h$ via combining all relevant features guided by coefficients $\mathbf{\alpha_{m,k}^{(h)}}$:
\begin{equation}\label{eqa::sum}
\mathbf{\widetilde{e}_m^{(h)}} = \sum_{k=1}^{M}\mathbf{\alpha_{m,k}^{(h)}}(\mathbf{W^{(h)}_{Value}}\mathbf{e_k}),
\end{equation}
where $\mathbf{W^{(h)}_{Value}} \in \mathbb{R}^{d'\times d}$.

Since $\mathbf{\widetilde{e}_m^{(h)}}\in \mathbb{R}^{d'}$ is a combination of feature $m$ and its relevant features (under head $h$), it represents a new combinatorial feature learned by our method. Furthermore, a feature is also likely to be involved in different combinatorial features, and we achieve this by using multiple heads, which create different subspaces and learn distinct feature interactions separately. We collect combinatorial features learned in all subspaces as follows:

\begin{figure}
\includegraphics[width=0.85\linewidth]{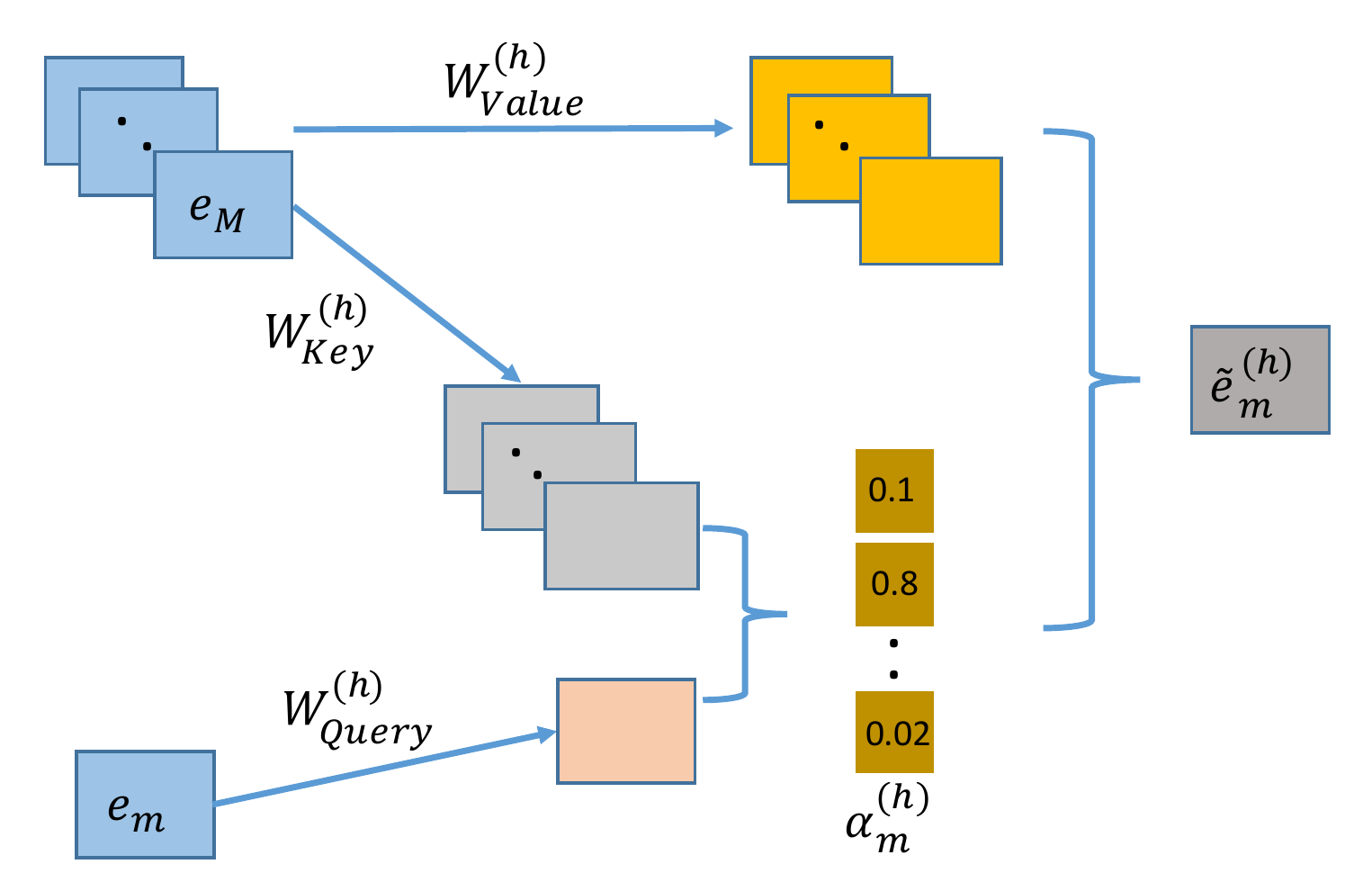}
\caption{The architecture of interacting layer. Combinatorial features are conditioned on attention weights, i.e., $\mathbf{\alpha_m^{(h)}}$.}\label{fig::attention}
\end{figure}

\begin{equation}
\mathbf{\widetilde{e}_m} = \mathbf{\widetilde{e}_m^{(1)}} \oplus\mathbf{\widetilde{e}_m^{(2)}}\oplus\cdot\cdot\cdot\oplus\mathbf{\widetilde{e}_m^{(H)}},
\end{equation}
where $\oplus$ is the concatenation operator, and H is the number of total heads. 

To preserve previously learned combinatorial features, including raw individual (i.e., first-order) features, we add standard residual connections in our network.
Formally,
\begin{equation}\label{eqa::res}
\mathbf{e_m^{Res}} = \textit{ReLU}(\mathbf{\widetilde{e}_m} + \mathbf{W_{Res}}\mathbf{e_m}),
\end{equation}
where $\mathbf{W_{Res}}\in \mathbb{R}^{d'H\times d}$ is the projection matrix in case of dimension mismatching~\cite{he2016deep}, and $\textit{ReLU}(z)=max(0,z)$ is a non-linear activation function.

With such an interacting layer, the representation of each feature $\mathbf{e_m}$ will be updated into a new feature representation $\mathbf{e_m^{Res}}$, which is a representation of high-order features. We can stack multiple such layers with the output of the previous interacting layer as the input of the next interacting layer. By doing this, we can model arbitrary-order combinatorial features.

 
\subsection{Output Layer}
The output of the interacting layer is a set of feature vectors $\{\mathbf{e_m^{Res}}\}_{m=1}^{M}$, which includes raw individual features reserved by residual block and combinatorial features learned via the multi-head self-attention mechanism. For final CTR prediction, we simply concatenate all of them and then apply a non-linear projection as follows:
\begin{equation}\label{eqa::ctr}
\hat{y}=\sigma(\mathbf{w^T}(\mathbf{e_1^{Res}}\oplus \mathbf{e_2^{Res}}\oplus\cdot\cdot\cdot\oplus \mathbf{e_M^{Res})}+b),
\end{equation}
where $\mathbf{w}\in \mathbb{R}^{d'HM}$ is a column projection vector which linearly combines concatenated features, $b$ is the bias, and $\sigma(x)=1/(1+e^{-x})$ transforms the values to users clicking probabilities.

\subsection{Training}
Our loss function is \textit{Log loss}, 
which is defined as follows:
\begin{equation}\label{eqa::loss}
Logloss = -\frac{1}{N}\sum^{N}_{j=1}(y_j \log(\hat{y}_j)+(1-y_j)\log(1-\hat{y}_j)),
\end{equation}
where $y_j$ and $\hat{y}_j$ are ground truth of user clicks and estimated CTR respectively, $j$ indexes the training samples, and $N$ is the total number of training samples. The parameters to learn in our model are \{$\mathbf{V_i}$, $\mathbf{v_m}$, $\mathbf{W^{(h)}_{Query}}$, $\mathbf{W^{(h)}_{Key}}$, $\mathbf{W^{(h)}_{Value}}$, $\mathbf{W_{Res}}$, $\mathbf{w}$, $b$\}, which are updated via minimizing the total \textit{Logloss} using gradient descent.

\subsection{Analysis Of AutoInt}
\smallskip
\xhdr{Modeling Arbitrary Order Combinatorial Features} Given feature interaction operator defined by Equation~\ref{eqa::align} - \ref{eqa::res}, we now analyze how low-order and high-order combinatorial features are modeled in our proposed model. 

For simplicity, let's assume there are four feature fields (i.e., $M$=4) denoted by $x_1$, $x_2$, $x_3$ and $x_4$ respectively. Within the first interacting layer, each individual feature interacts with any other features through attention mechanism (i.e. Equation~\ref{eqa::align}) and therefore a set of second-order feature combinations such as $g(x_1, x_2)$, $g(x_2, x_3)$ and $g(x_3, x_4)$ are captured with distinct correlation weights, where the non-additive property of interaction function $g(\cdot)$ (in DEFINITION 2) can be ensured by the non-linearity of activation function $\textit{ReLU}(\cdot)$. Ideally, combinatorial features that involve $x_1$ can be encoded into the updated representation of the first feature field $\mathbf{e_1^{Res}}$.
As the same can be derived for other feature fields, all second-order feature interactions can be encoded in the output of the first interacting layer, where attention weights distill useful feature combinations.   

Next, we prove that higher-order feature interactions can be modeled within the second interacting layer. Given the representation of the first feature field $\mathbf{e_1^{Res}}$ and the representation of the third feature field $\mathbf{e_3^{Res}}$ generated by the first interacting layer, third-order combinatorial features that involve $x_1$, $x_2$ and $x_3$ can be modeled by allowing $\mathbf{e_1^{Res}}$ to attend on $\mathbf{e_3^{Res}}$ because $\mathbf{e_1^{Res}}$ contains the interaction $g(x_1, x_2)$ and $\mathbf{e_3^{Res}}$ contains the individual feature $x_3$ (from residual connection). Moreover, the maximum order of combinatorial features grows exponentially with respect to the number of interacting layers. For example, fourth-order feature interaction $g(x_1, x_2, x_3, x_4)$ can be captured by the combination of $\mathbf{e_1^{Res}}$ and $\mathbf{e_3^{Res}}$, which contain the second-order interactions $g(x_1, x_2)$ and $g(x_3, x_4)$ respectively.
Therefore a few interacting layers will suffice to model high-order feature interactions.

Based on above analysis, we can see that AutoInt learns feature interactions with attention mechanism in a hierarchical manner, i.e., from low-order to high-order, and all low-order feature interactions are carried by residual connections. This is promising and reasonable because learning hierarchical representation has proven quite effective in computer vision and speech processing with deep neural networks~\cite{lee2011unsupervised,bengio2013representation}.

\smallskip
\xhdr{Space Complexity}
The embedding layer, which is a shared component in neural network-based methods~\cite{lian2018xdeepfm, guo2017deepfm, shan2016deep}, contains $nd$ parameters, where $n$ is the dimension of sparse representation of input feature and $d$ is the embedding size.
As an interacting layer contains following weight matrices: \{$\mathbf{W^{(h)}_{Query}}, \mathbf{W^{(h)}_{Key}}, \mathbf{W^{(h)}_{Value}}, \mathbf{W_{Res}}$\}, the number of parameters in an $L$-layer network is $L\times (3dd'+d'Hd)$, which is independent of the number of feature fields $M$. Finally, there are $d'HM+1$ parameters in the output layer. As far as interacting layers are concerned, the space complexity is $O(Ldd'H)$. Note that $H$ and $d'$ are usually small (e.g., $H=2 \text{ and } d'=32$ in our experiments), which makes the interacting layer memory-efficient.

\smallskip
\xhdr{Time Complexity} Within each interacting layer, the computation cost is two-fold. First, calculating attention weights for one head takes $O(Mdd' + M^2d')$ time. Afterwards, forming combinatorial features under one head also takes $O(Mdd' + M^2d')$ time. Because we have $H$ heads, it takes $O(MHd^\prime(M+d))$ time altogether. It is therefore efficient because $H, d$ and $d'$ are usually small. We provide running time of AutoInt in Section~\ref{sec::result}.


%% file: experiment.tex
\section{Experiment}
In this section, we move forward to evaluate the effectiveness of our proposed approach. We aim to answer the following questions:


\begin{itemize}[leftmargin=6ex,labelsep=1ex]
\raggedright
	\item[\textbf{RQ1}] How does our proposed AutoInt perform on the problem of CTR prediction? Is it efficient for large-scale sparse and high-dimensional data?
	\item[\textbf{RQ2}] What are the influences of different model configurations?
	\item[\textbf{RQ3}] What are the dependency structures between different features? Is our proposed model explainable?
   	\item[\textbf{RQ4}] Will integrating implicit feature interactions further improve the performance?
\end{itemize}
We first describe the experimental settings before answering these questions.

\subsection{Experiment Setup}

\subsubsection{Data Sets}\label{subsubsec::data}
We use four public real-world data sets. The statistics of the data sets are summarized in Table~\ref{tab::dataset}.
\textbf{Criteo}\footnote{https://www.kaggle.com/c/criteo-display-ad-challenge} This is a benchmark dataset for CTR prediction, which has 45 million users' clicking records on displayed ads. It contains 26 categorical feature fields and 13 numerical feature fields. 
\textbf{Avazu}\footnote{https://www.kaggle.com/c/avazu-ctr-prediction} This dataset contains users' mobile behaviors including whether a displayed mobile ad is clicked by a user or not. It has 23 feature fields spanning from user/device features to ad attributes.  
\textbf{KDD12}\footnote{https://www.kaggle.com/c/kddcup2012-track2} This data set was released by KDDCup 2012, which originally aimed to predict the number of clicks. Since our work focuses on CTR prediction rather than the exact number of clicks, we treat this problem as a binary classification problem (1 for clicks>0, 0 for without click), which is similar to FFM~\cite{juan2016field}.
\textbf{MovieLens-1M}\footnote{https://grouplens.org/datasets/movielens/} This dataset contains users' ratings on movies. During binarization, we treat samples with a rating less than 3 as negative samples because a low score indicates that the user does not like the movie. We treat samples with a rating greater than 3 as positive samples and remove neutral samples, i.e., a rating equal to 3.

\noindent\textbf{Data Preparation} First, we remove the infrequent features (appearing in less than \textit{threshold} instances) and treat them as a single feature ``<unknown>'', where \textit{threshold} is set to \{10, 5, 10\} for Criteo, Avazu and KDD12 data sets respectively.
Second, since numerical features may have large variance and hurt machine learning algorithms, we normalize numerical values by transforming a value $z$ to $log^2(z)$ if $z > 2$, which is proposed by the winner of Criteo Competition\footnote{\url{https://www.csie.ntu.edu.tw/~r01922136/kaggle-2014-criteo.pdf}}. Third, we randomly select 80\% of all samples for training and randomly split the rest into validation and test sets of equal size.

\begin{table}
\centering\caption{Statistics of evaluation data sets.}
\begin{tabular}{cccc} 
\toprule
Data & \#Samples & \#Fields & \#Features (Sparse)  \\
\midrule
Criteo & 45,840,617 & 39 & 998,960 \\
Avazu & 40,428,967 & 23 & 1,544,488 \\
KDD12 & 149,639,105 & 13 & 6,019,086\\
MovieLens-1M & 739,012 & 7 & 3,529 \\
\bottomrule
\end{tabular}\label{tab::dataset}
\end{table}

\begin{table*}
\begin{threeparttable}
\centering\caption{Effectiveness Comparison of Different Algorithms. We highlight that our proposed model almost outperforms all baselines across four data sets and both metrics. Further analysis is provided in Section~\ref{sec::result}.}\label{tab::results}
\tabcolsep=0.24cm
\begin{tabular}{llcccccccc}
\toprule
\multirow{2}{*}{Model Class} & \multirow{2}{*}{Model} & \multicolumn{2}{c}{Criteo} & \multicolumn{2}{c}{Avazu} & \multicolumn{2}{c}{KDD12} & \multicolumn{2}{c}{MovieLens-1M} \\
& & AUC & Logloss & AUC & Logloss & AUC & Logloss & AUC & Logloss \\
\midrule
\multirow{1}{*}{First-order} & LR & 0.7820 & 0.4695 & 0.7560 & 0.3964 & 0.7361 & 0.1684 & 0.7716 & 0.4424 \\
 \midrule
\multirow{2}{*}{Second-order} & FM~\cite{rendle2010factorization} & 0.7836 & 0.4700 & 0.7706 & 0.3856 & 0.7759 & 0.1573 & 0.8252 & 0.3998 \\ 
& AFM\cite{xiao2017attentional} & 0.7938 & 0.4584 & 0.7718 & 0.3854 & 0.7659 & 0.1591 & 0.8227 & 0.4048 \\
\midrule
\multirow{6}{*}{High-order} & DeepCrossing~\cite{shan2016deep} & 0.8009 &0.4513 & 0.7643 & 0.3889 & 0.7715 & 0.1591 & 0.8448 & 0.3814 \\
& NFM~\cite{he2017neural} & 0.7957 & 0.4562 & 0.7708 & 0.3864 & 0.7515 & 0.1631 & 0.8357 & 0.3883 \\
& CrossNet~\cite{wang2017deep} & 0.7907 & 0.4591 & 0.7667 & 0.3868 & 0.7773 & 0.1572 & 0.7968 & 0.4266 \\
& CIN~\cite{lian2018xdeepfm} & 0.8009 & 0.4517 & \textbf{0.7758} & 0.3829 & 0.7799 & 0.1566 & 0.8286 & 0.4108 \\
& HOFM~\cite{blondel2016higher} & 0.8005 & 0.4508 & 0.7701 & 0.3854 & 0.7707 & 0.1586 & 0.8304 & 0.4013 \\
& AutoInt (ours)  & \textbf{0.8061**} & \textbf{0.4455**} & 0.7752 & \textbf{0.3824} & \textbf{0.7883**} & \textbf{0.1546**} & \textbf{0.8456*} & \textbf{0.3797**}  \\
\bottomrule
\end{tabular}
    \begin{tablenotes}
    \centering
      \small
      \item AutoInt outperforms the strongest baseline w.r.t. Criteo, KDD12 and MovieLens-1M data at the: \textbf{**} 0.01 and \textbf{*} 0.05 level, unpaired t-test.
    \end{tablenotes}
\end{threeparttable}
\end{table*}

\subsubsection{Evaluation Metrics}
We use two popular metrics to evaluate the performance of all methods.

\textbf{AUC} Area Under the ROC Curve (AUC) measures the probability that a CTR predictor will assign a higher score to a randomly chosen positive item than a randomly chosen negative item. A higher AUC indicates a better performance.

\textbf{Logloss} Since all models attempt to minimize the \textit{Logloss} defined by Equation~\ref{eqa::loss}, we use it as a straightforward metric.

It is noticeable that a slightly higher AUC or lower \textit{Logloss} at \textit{\textbf{0.001-level}} is regarded significant for CTR prediction task, which has also been pointed out in existing works~\cite{cheng2016wide,guo2017deepfm,wang2017deep}.

\subsubsection{Competing Models} We compare the proposed approach with three classes of previous models. (A) the linear approach that only uses individual features. (B) factorization machines-based methods that take into account second-order combinatorial features. (C) techniques that can capture high-order feature interactions. We associate the model classes with model names accordingly.

\textbf{LR} (A). LR only models the linear combination of raw features.  

\textbf{FM}~\cite{rendle2010factorization} (B). FM uses factorization techniques to model second-order feature interactions.

\textbf{AFM}~\cite{xiao2017attentional} (B). AFM is one of the state-of-the-art models that capture second-order feature interactions. It extends FM by using attention mechanism to distinguish the different importance of second-order combinatorial features.

\textbf{DeepCrossing}~\cite{shan2016deep} (C). DeepCrossing utilizes deep fully-connected neural networks with residual connections to learn non-linear feature interactions in an implicit fashion.

\textbf{NFM}~\cite{he2017neural} (C). NFM stacks deep neural networks on top of second-order feature interaction layer. High-order feature interactions are implicitly captured by the nonlinearity of neural networks.

\textbf{CrossNet}~\cite{wang2017deep} (C). Cross Network, which is the core of Deep\&Cross model, takes outer product of concatenated feature vector at the bit-wise level to model feature interactions explicitly.

\textbf{CIN}~\cite{lian2018xdeepfm} (C). Compressed Interaction Network, which is the core of xDeepFM model, takes outer product of stacked feature matrix at vector-wise level.

\textbf{HOFM}~\cite{blondel2016higher} (C). HOFM proposes efficient kernel-based algorithms for training high-order factorization machines. Follow settings in~\citeauthor{blondel2016higher}~\cite{blondel2016higher} and \citeauthor{he2017neural}~\cite{he2017neural}, we build a third-order factorization machine using public implementation.

We will compare with the full models of CrossNet and CIN, i.e., Deep\&Cross and xDeepFM, under the setting of joint training with plain DNN later (i.e., Section~\ref{sec::joint}).

\begin{figure*}
\centering
  \begin{subfigure}[b]{0.24\linewidth}
    \includegraphics[width=\linewidth]{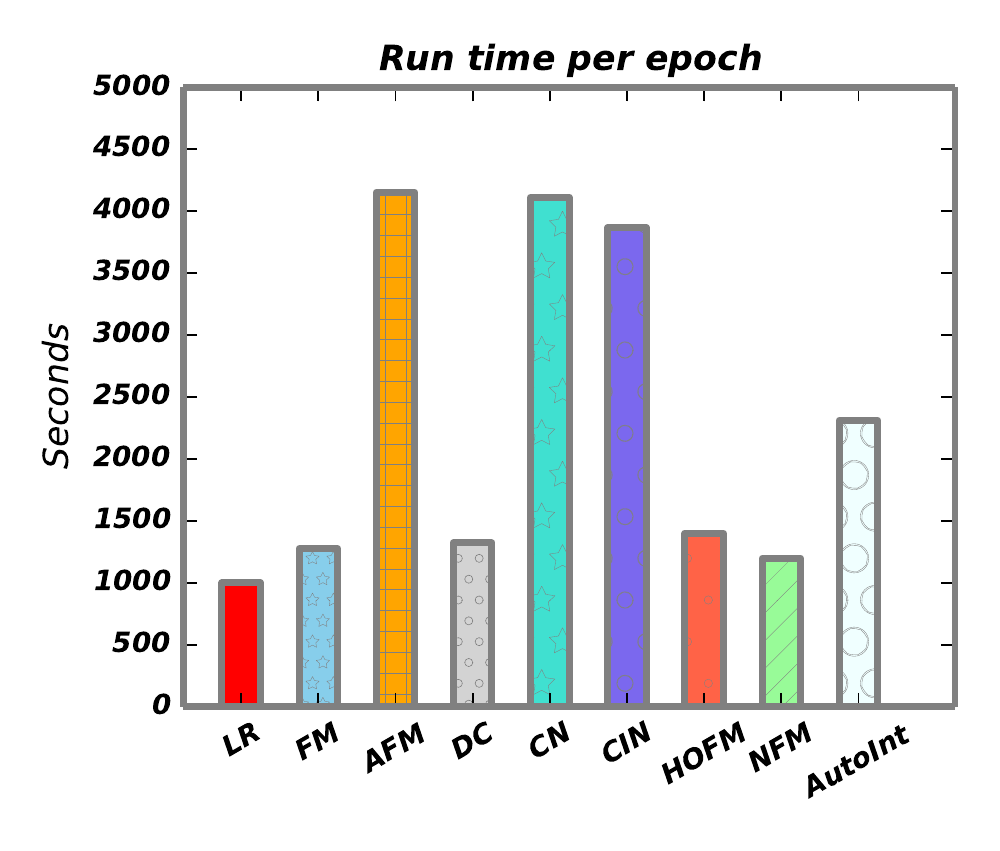}
    \caption{Criteo}
  \end{subfigure}
  \begin{subfigure}[b]{0.24\linewidth}
    \includegraphics[width=\linewidth]{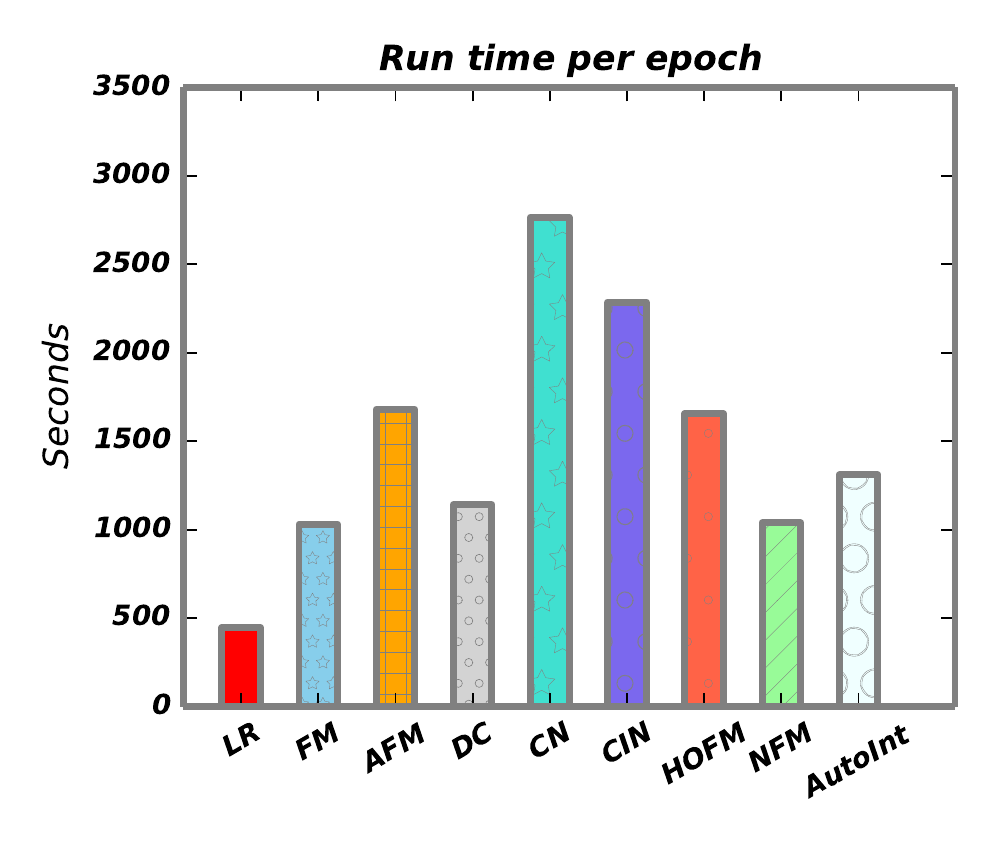}
    \caption{Avazu}
  \end{subfigure}
  \begin{subfigure}[b]{0.24\linewidth}
    \includegraphics[width=\linewidth]{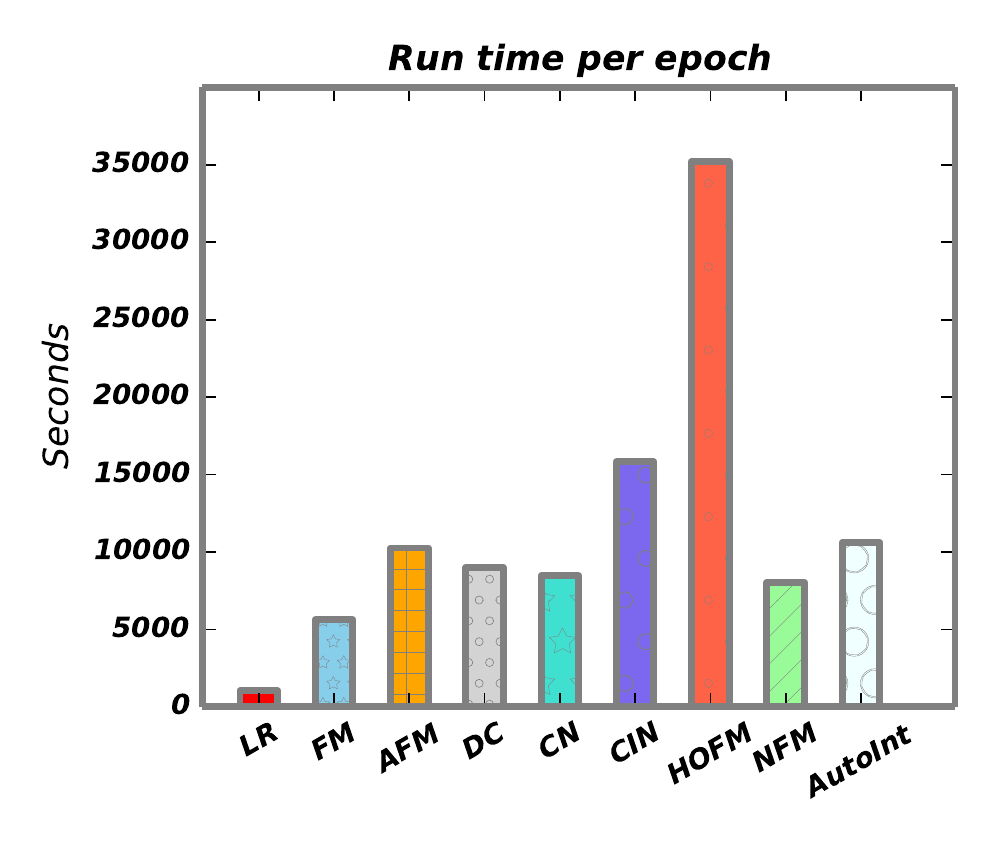}
    \caption{KDD12}
  \end{subfigure}
  \begin{subfigure}[b]{0.24\linewidth}
    \includegraphics[width=\linewidth]{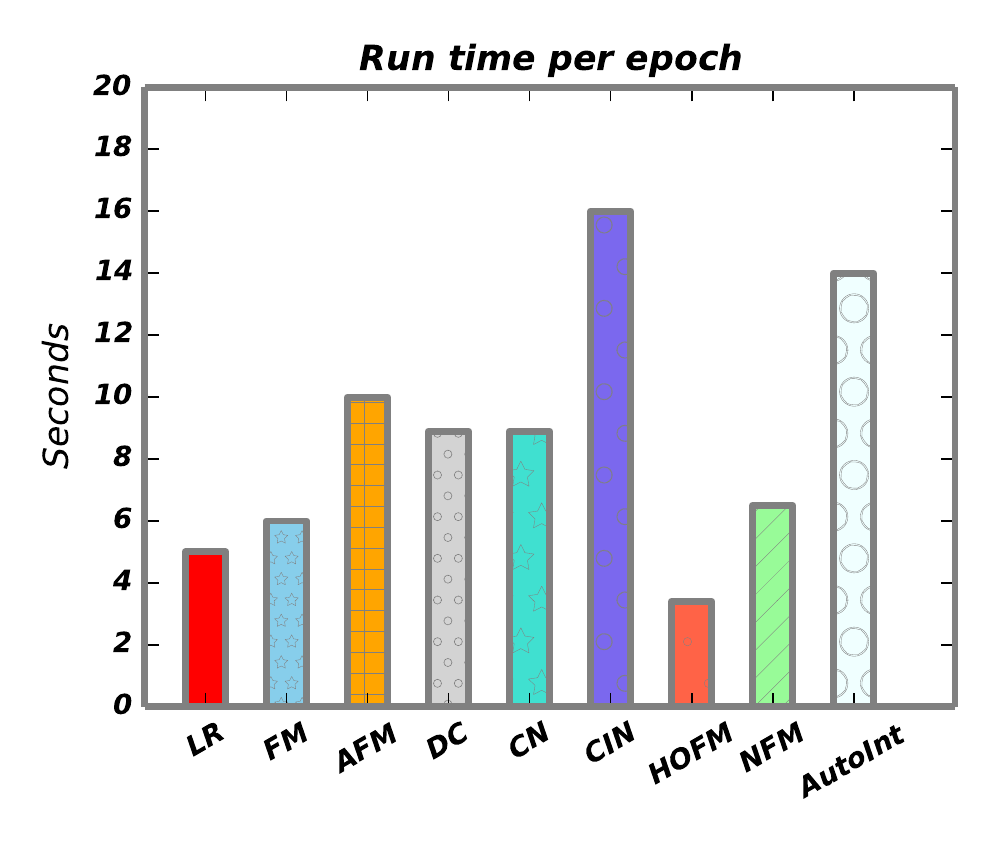}
    \caption{MovieLens-1M}
  \end{subfigure}
  \caption{Efficiency Comparison of Different Algorithms in terms of \textit{Run Time}. ``DC'' and ``CN'' are DeepCrossing and CrossNet for short, respectively. Since HOFM cannot be fit on one GPU card for the KDD12 dataset, extra communication cost makes it most time-consuming. Further analysis is presented in Section~\ref{sec::result}.}\label{fig::run_time}
  \vspace{-8pt}
\end{figure*}

\subsubsection{Implementation Details}\label{sec::impl}
All methods are implemented in TensorFlow\cite{abadi2016tensorflow}. For AutoInt and all baseline methods, we empirically set embedding dimension $d$ to 16 and batch size to 1024. AutoInt has three interacting layers and the number of hidden units $d'$ is 32 in default setting. Within each interacting layer, the number of attention head is two\footnote{We also tried different number of attention heads. The performance of using one head is inferior to that of two heads, and the improvement of further increasing head number is not significant.}. To prevent overfitting, we use grid search to select dropout rate~\cite{srivastava2014dropout} from \{0.1 - 0.9\} for MovieLens-1M data set, and we found dropout is not necessary for other three large data sets.
For baseline methods, we use one hidden layer of size 200 on top of Bi-Interaction layer for NFM as recommended by their paper. For CN and CIN, we use three interaction layers following AutoInt.  DeepCrossing has four feed-forward layers and the number of hidden units is 100, because it performs poorly when using three neural layers. Once all network structures are fixed, we also apply grid search to baseline methods for optimal hype-parameters. 
Finally, we use Adam~\cite{kingma2014adam} to optimize all deep neural network-based models.

\subsection{Quantitative Results (RQ1)}\label{sec::result}

\noindent\textbf{Evaluation of Effectiveness}\\
We summarize the results averaged over 10 different runs into Table~\ref{tab::results}.
We have the following observations: 
(1) FM and AFM, which explore second-order feature interactions, consistently outperform LR by a large margin on all datasets, which indicates that individual features are insufficient in CTR prediction.
(2) An interesting observation is the inferiority of some models which capture high-order feature interactions. For example, although DeepCrossing and NFM use the deep neural network as a core component to learning high-order feature interactions, they do not guarantee improvement over FM and AFM. This may attribute to the fact that they learn feature interactions in an implicit fashion. On the contrary, CIN does it explicitly and outperforms low-order models consistently.
(3) HOFM significantly outperforms FM on Criteo and MovieLens-1M datasets, which indicates that modeling third-order feature interactions can be beneficial to prediction performance.
(4) AutoInt achieves the best performance overall baseline methods on three of four real-world data sets. On Avazu data set, CIN performs a little better than AutoInt in AUC evaluation, but we get lower \textit{Logloss}. Note that our proposed AutoInt shares the same structures as DeepCrossing except the feature interacting layer, which indicates using the attention mechanism to learn explicit combinatorial features is crucial. \\

\noindent\textbf{Evaluation of Model Efficiency}\\
We present the runtime results of different algorithms on four data sets in Figure~\ref{fig::run_time}. Unsurprisingly, LR is the most efficient algorithm due to its simplicity. FM and NFM perform similarly in terms of runtime because NFM only stacks a single feed-forward hidden layer on top of the second-order interaction layer. Among all listed methods, CIN, which achieves the best performance for prediction among all the baselines, is much more time-consuming due to its complicated crossing layer. This may make it impractical in the industrial scenarios. Note that AutoInt is sufficiently efficient, which is comparable to the efficient algorithms DeepCrossing and NFM.

We also compare the sizes of different models (i.e., the number of parameters) as another criterion for efficiency evaluation. As shown in Table~\ref{tab::param}, comparing to the best model CIN in the baseline models, the number of parameters in AutoInt is much smaller. 


To summarize, our proposed AutoInt achieves the best performance among all the compared models. Compared to the most competitive baseline model CIN, AutoInt requires much fewer parameters and is much more efficient during online inference. 


\begin{table}
\centering\caption{Efficiency Comparison of Different Algorithms in terms of \textit{Model Size} on Criteo data set. ``DC'' and ``CN'' are DeepCrossing and CrossNet for short, respectively. The counted parameters exclude the embedding layer.}
\begin{tabularx}{\linewidth}{cccccc}
\toprule
Model & DC & CN & CIN & NFM & AutoInt \\
\midrule
\#Params & $1.6\times10^5$ & $3\times10^3$ & $1.9\times10^6$ & $4\times10^3$ & $3.9\times10^4$ \\
\bottomrule
\end{tabularx}
\vspace{-6pt}
\label{tab::param}
\end{table}

\begin{table}
\centering\caption{Ablation study comparing the performance of AutoInt with and without residual connections. AutoInt$_\text{{w/}}$ is the complete model while the AutoInt$_\text{{w/o}}$ is the model without residual connection.}
\begin{tabularx}{1.0\linewidth}{l>{\centering\arraybackslash}l>{\centering\arraybackslash}X>{\centering\arraybackslash}X}
\toprule
Data Sets & Models & AUC & Logloss \\
\midrule
\multirow{2}{*}{Criteo} & AutoInt$_\text{{w/}}$ & 0.8061 & 0.4454 \\
 & AutoInt$_\text{{w/o}}$ & 0.8033 & 0.4478 \\
\midrule
\multirow{2}{*}{Avazu} & AutoInt$_\text{{w/}}$ & 0.7752 & 0.3823 \\
 & AutoInt$_\text{{w/o}}$ & 0.7729 & 0.3836 \\
 \midrule
\multirow{2}{*}{KDD12} & AutoInt$_\text{{w/}}$ & 0.7888 & 0.1545 \\
 & AutoInt$_\text{{w/o}}$ & 0.7831 & 0.1557 \\
\midrule
\multirow{2}{*}{MovieLens-1M} & AutoInt$_\text{{w/}}$ & 0.8460 & 0.3784 \\ 
 & AutoInt$_\text{{w/o}}$ & 0.8299 & 0.3959 \\
\bottomrule
\end{tabularx}
\vspace{-6pt}
\label{tab::residual}
\end{table}

\subsection{Analysis (RQ2)}
To further validate and gain deep insights into the proposed model, we conduct ablation study and compare several variants of AutoInt. 

\subsubsection{Influence of Residual Structure}
The standard AutoInt utilizes residual connections, which carry through all learned combinatorial features and therefore allow modeling very high-order combinations. To justify the contribution of residual units, we tease apart them from our standard model and keep other structures as they are. As presented in Table~\ref{tab::residual}, we observe that the performance decrease on all datasets if residual connections are removed. Specifically, the full model outperforms the variant by a large margin on the KDD12 and MovieLens-1M data, which indicates residual connections are crucial to model high-order feature interactions in our proposed method.

\subsubsection{Influence of Network Depths}
Our model learns high-order feature combinations by stacking multiple interacting layers (introduced in Section~\ref{sec::model}). Therefore, we are interested in how the performance change w.r.t. the number of interacting layers, i.e., the order of combinatorial features. 
Note that when there is no interacting layer (i.e., \textit{Number of layers} equals zero), our model takes the weighted sum of raw individual features as input, i.e., no combinatorial features are considered. 

\begin{figure}
\centering
\begin{subfigure}[b]{0.49\linewidth}
    \includegraphics[width=\linewidth]{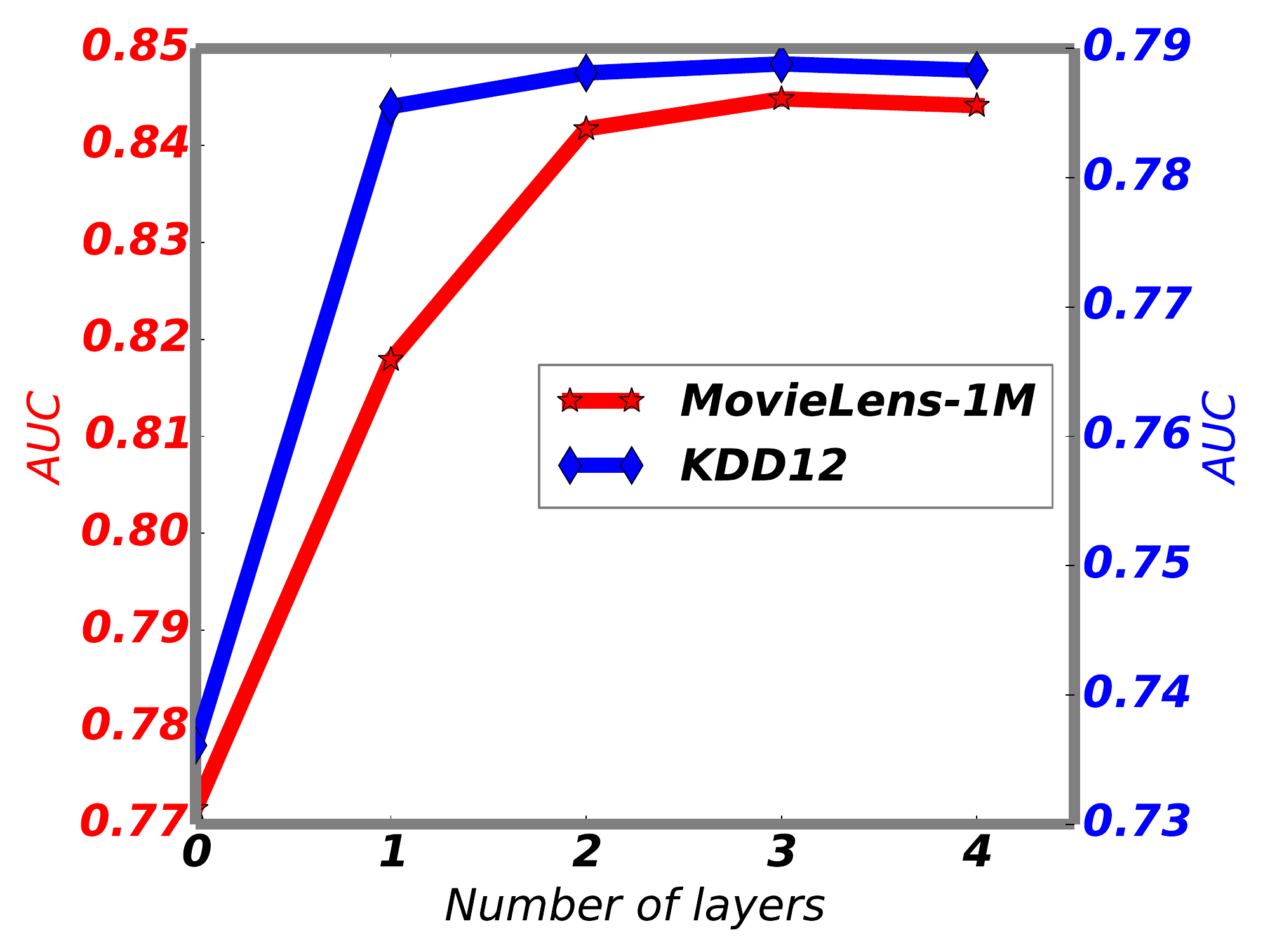}
    \caption{AUC}
  \end{subfigure}
  \begin{subfigure}[b]{0.49\linewidth}
    \includegraphics[width=\linewidth]{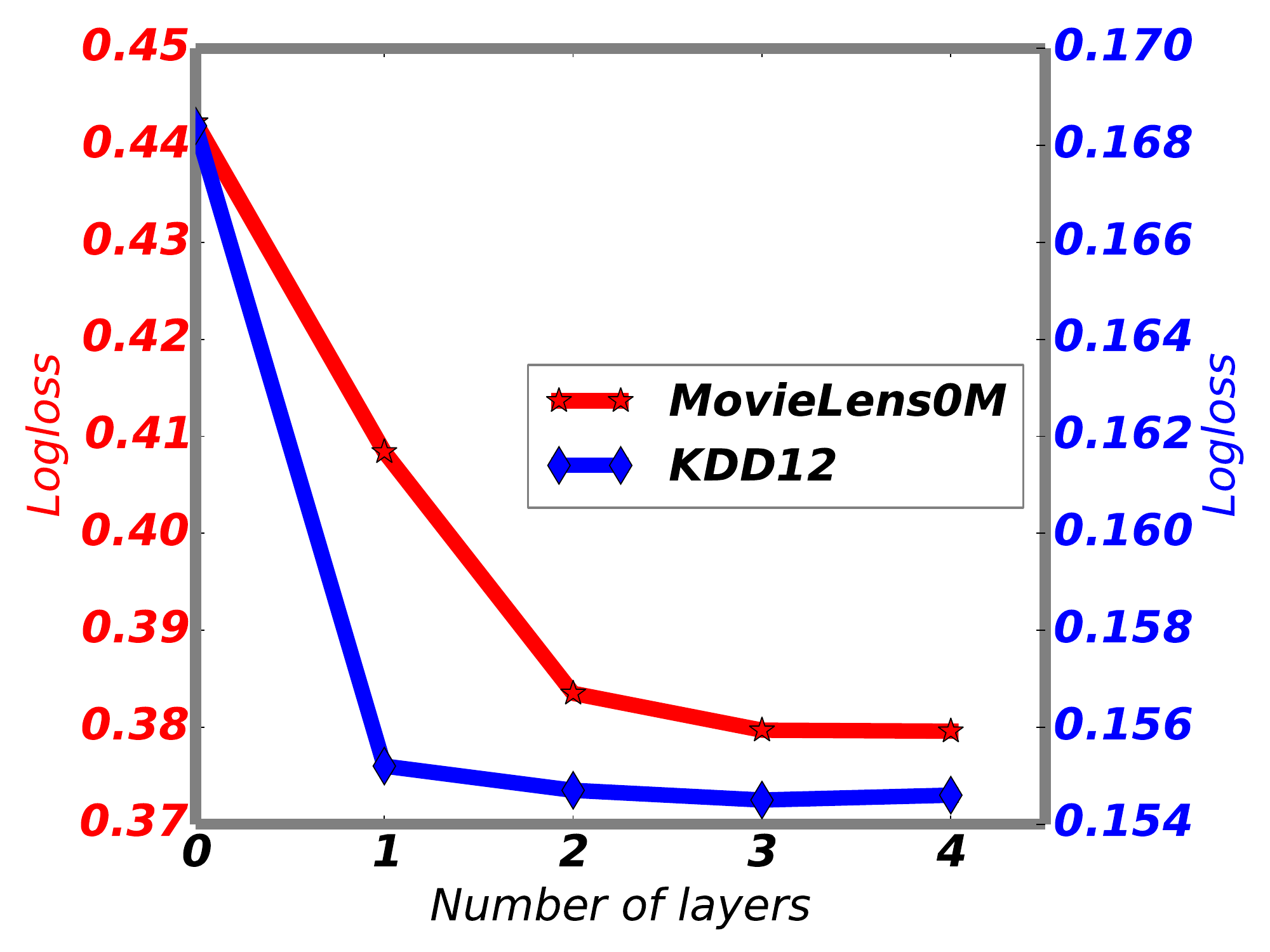}
    \caption{Logloss}
  \end{subfigure}
  \caption{Performance w.r.t. the number of interacting layers. Results on Criteo and Avazu data sets are similar and hence omitted.}
  \vspace{-10pt}
  \label{fig::layer}
\end{figure}

The results are summarized in Figure~\ref{fig::layer}. We can see that if one interacting layer is used, i.e., feature interactions are taken into account, the performance increase dramatically on both data sets, showing that combinatorial features are very informative for prediction. As the number of interacting layers further increases, i.e., higher-order combinatorial features are taken into account, the performance of the model further increases. When the number of layers reaches three, the performance becomes stable, showing that adding extremely high-order features are not informative for prediction. 


\begin{table*}
\begin{threeparttable}
\centering\caption{Results of Integrating Implicit Feature Interactions. We indicate the base model behind each method. The last two columns are average changes of AUC and \textit{Logloss} compared to corresponding base models (``+'': increase, ``-'': decrease).}\label{tab::ensemble}
\begin{tabular}{lcccccccc||cc}
\toprule
\multirow{2}{*}{Model} & \multicolumn{2}{c}{Criteo} & \multicolumn{2}{c}{Avazu} & \multicolumn{2}{c}{KDD12} & \multicolumn{2}{c}{MovieLens-1M} & \multicolumn{2}{c}{Avg. Changes}\\
& AUC & Logloss & AUC & Logloss & AUC & Logloss & AUC & Logloss & AUC & Logloss\\
\midrule
Wide\&Deep (LR) & 0.8026 & 0.4494 & 0.7749 & 0.3824 & 0.7549 & 0.1619 & 0.8300 & 0.3976 & +0.0292 & -0.0213 \\
DeepFM (FM) & 0.8066 & 0.4449 & 0.7751 & 0.3829 & 0.7867 & 0.1549 & 0.8437 & 0.3846 & +0.0142 & -0.0113\\
Deep\&Cross (CN) & 0.8067 & 0.4447 & 0.7731 & 0.3836 & 0.7872 & 0.1549 & 0.8446 & 0.3809 & +0.0200 & -0.0164 \\
xDeepFM (CIN) & 0.8070 & 0.4447 & 0.7770 & 0.3823 & 0.7820 & 0.1560 & 0.8463 & 0.3808 & +0.0068 & -0.0096 \\
AutoInt+ (ours) & \textbf{0.8083**} & \textbf{0.4434**} & \textbf{0.7774*} & \textbf{0.3811**} & \textbf{0.7898**} & \textbf{0.1543**} & \textbf{0.8488**} & \textbf{0.3753**} & +0.0023 & -0.0020 \\ 
\bottomrule
\end{tabular}
    \begin{tablenotes}
    \centering
      \small
      \item AutoInt+ outperforms the strongest baseline w.r.t. each data at the: \textbf{**} 0.01 and \textbf{*} 0.05 level, unpaired t-test.
    \end{tablenotes}
\end{threeparttable}
\vspace{-5pt}
\end{table*}

\begin{figure}
\centering
\begin{subfigure}[b]{0.49\linewidth}
    \includegraphics[width=\linewidth]{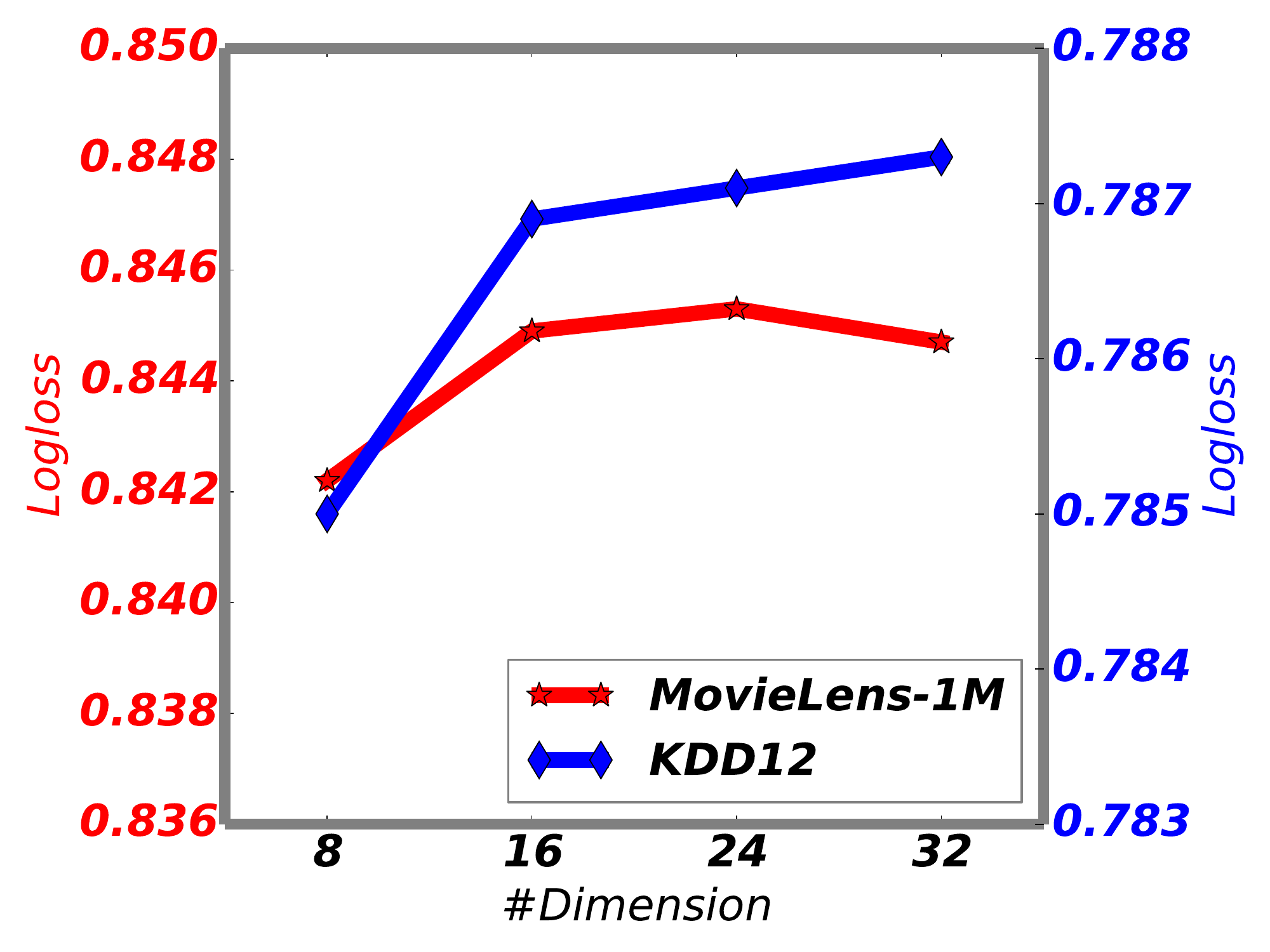}
    \caption{AUC}
  \end{subfigure}
  \begin{subfigure}[b]{0.49\linewidth}
    \includegraphics[width=\linewidth]{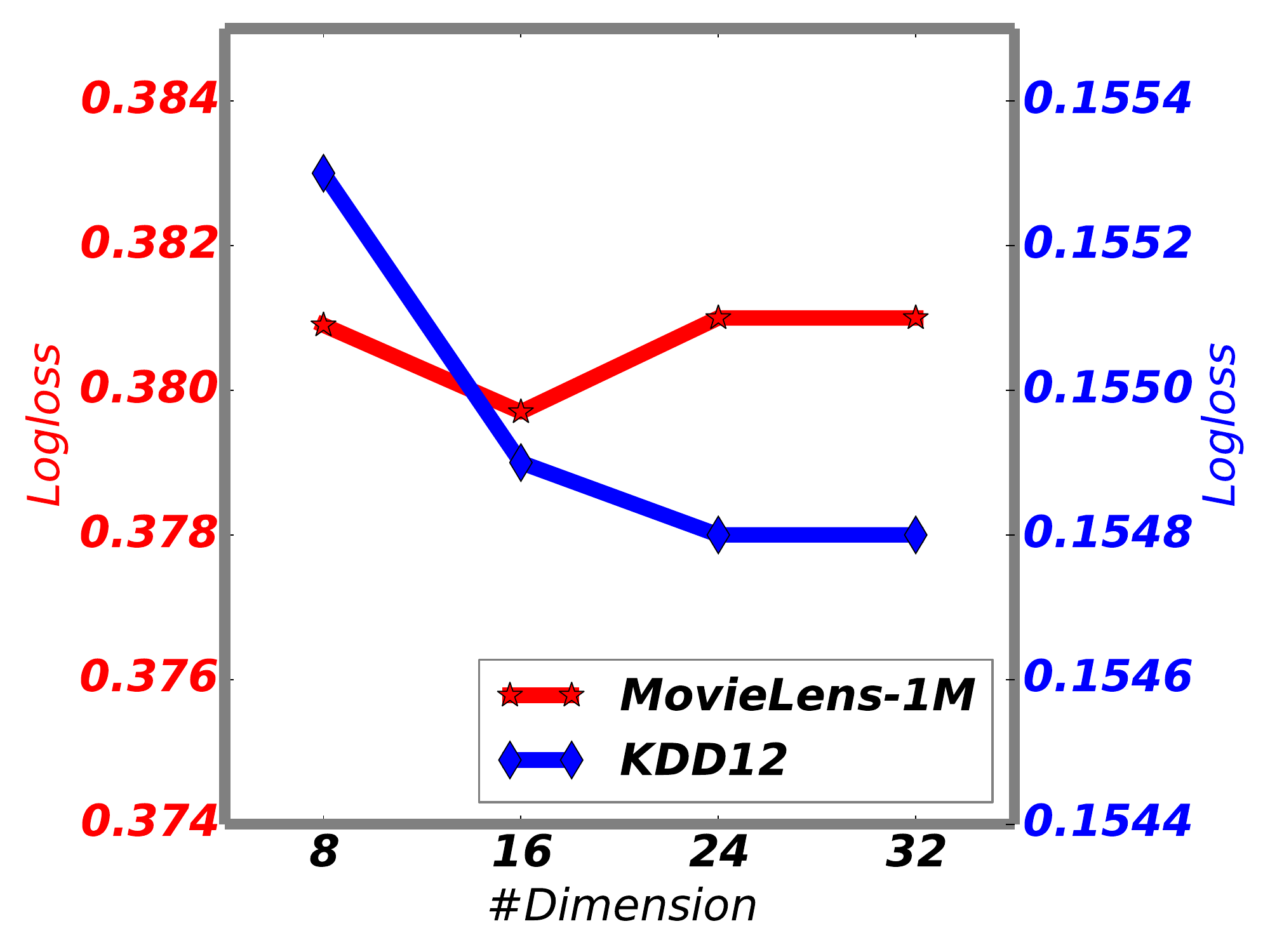}
    \caption{Logloss}
  \end{subfigure}
  \caption{Performance w.r.t. number of embedding dimensions. Results on Criteo and Avazu data sets are similar and hence omitted.}
  \vspace{-10pt}
  \label{fig::dim}
\end{figure}

\subsubsection{Influence of Different Dimensions}
Next, we investigate the performance w.r.t. the parameter $d$, which is the output dimension of the embedding layer. On the KDD12 dataset, we can see that the performance continuously increase as we increase the dimension size since larger models are used for prediction. The results are different on the MovieLens-1M dataset. When the dimension size reaches 24, the performance begins to decrease. The reason is that this data set is small, and the model is overfitted when too many parameters are used.



\begin{figure}
\centering
\begin{subfigure}[b]{0.495\linewidth}
    \includegraphics[width=\linewidth]{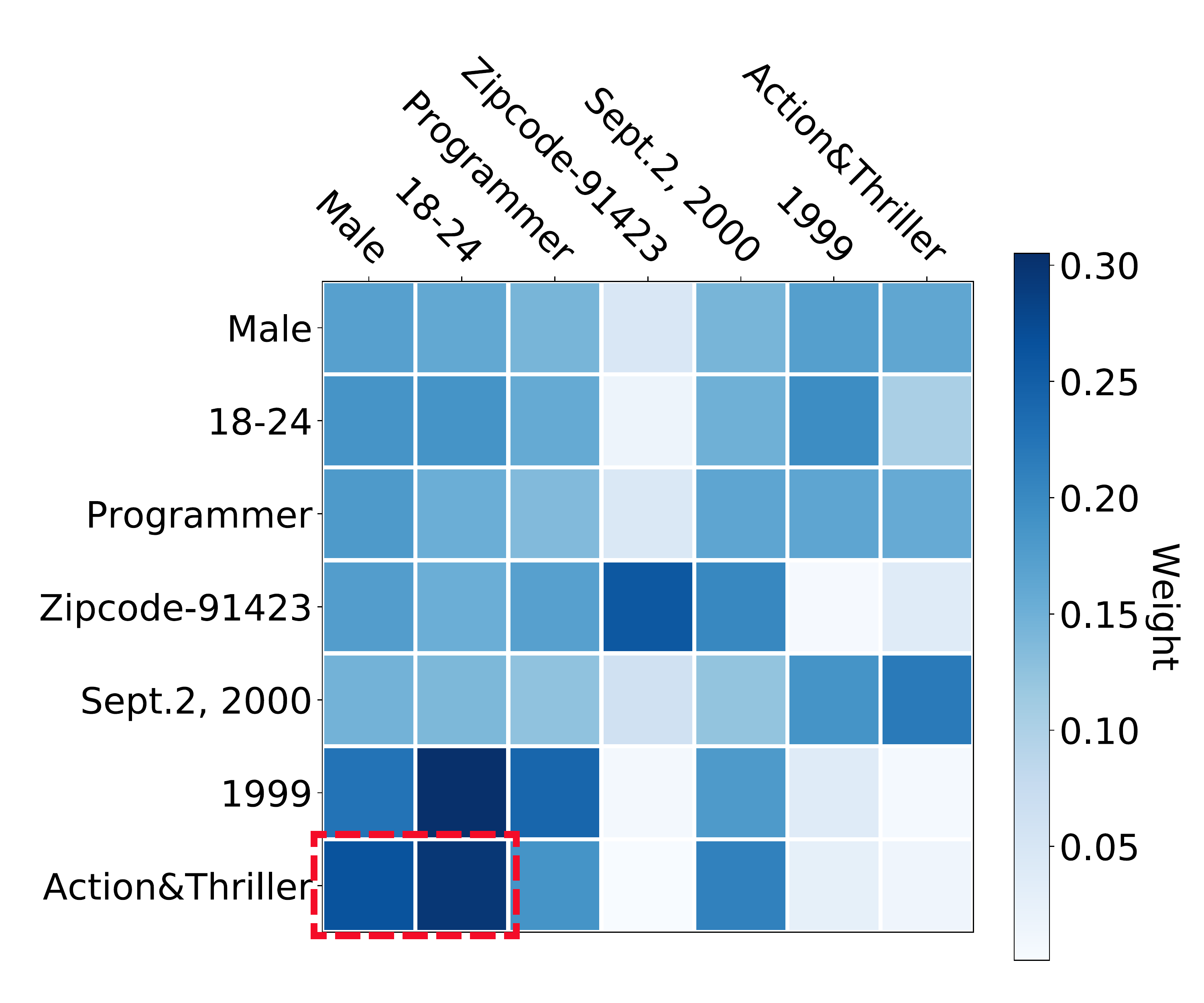}
    \caption{Label=1, Predicted CTR=0.89}
  \end{subfigure}
	\begin{subfigure}[b]{0.495\linewidth}
      \includegraphics[width=\linewidth]{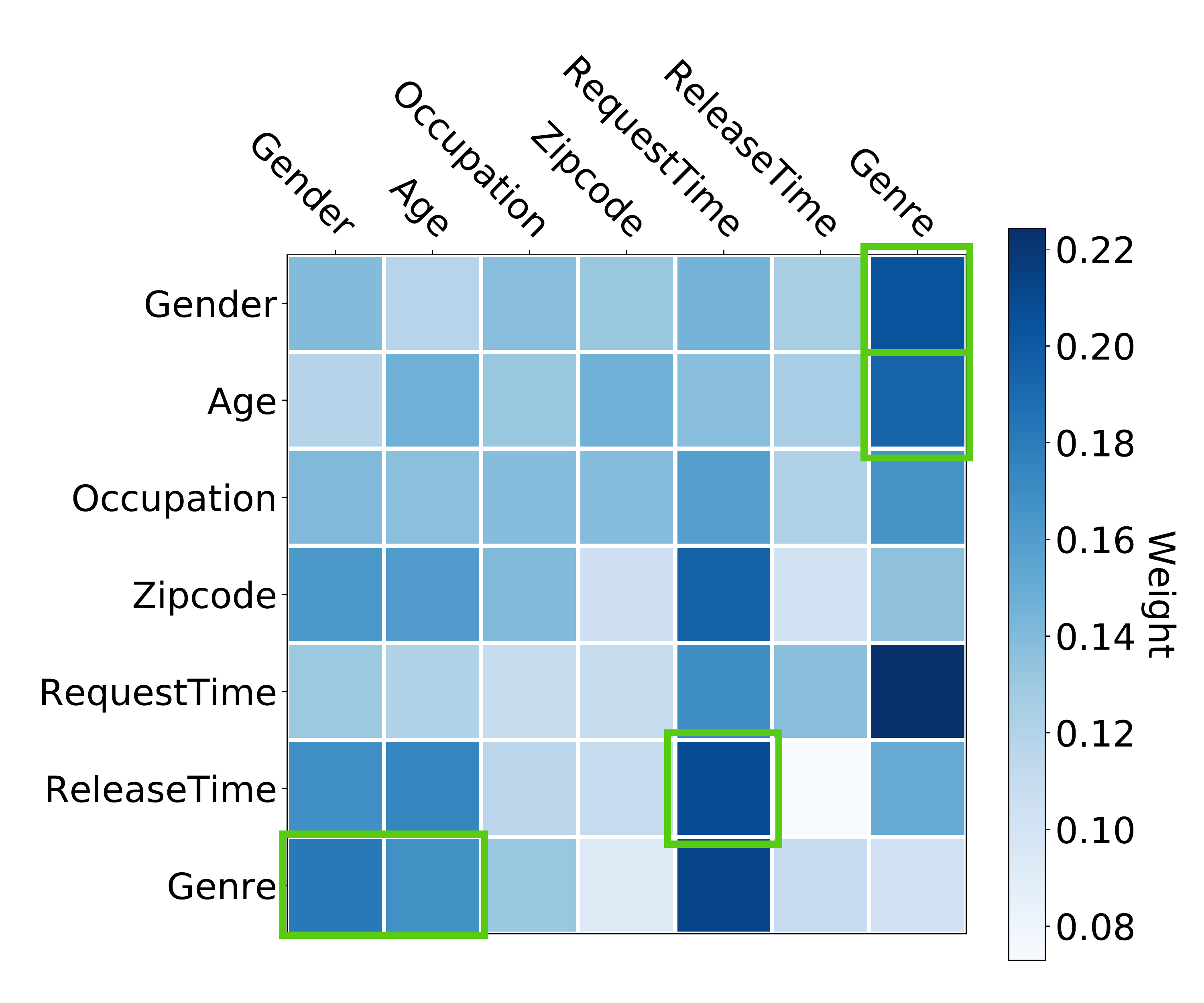}
      \caption{Overall feature interactions}
  	\end{subfigure}
  \caption{Heat maps of attention weights for both case- and global-level feature interactions on MovieLens-1M. The axises represent feature fields <\textit{Gender, Age, Occupation, Zipcode, RequestTime, RealeaseTime, Genre}>. We highlight some learned combinatorial features in rectangles.}
  \vspace{-11pt}
  \label{fig::case}
\end{figure}

\subsection{Explainable Recommendations (RQ3)}
A good recommender system can not only provide good recommendations but also offer good explainability. Therefore, in this part, we present how our AutoInt is able to explain the recommendation results. 
We take the MovieLens-1M dataset as an example. 


Let's look at a recommendation result suggested by our algorithm, i.e., a user likes an item. Figure~\ref{fig::case} (a) presents the correlations between different fields of input features, which are obtained by the attention score. We can see that AutoInt is able to identify the meaningful combinatorial feature <\textit{Gender=Male, Age=[18-24), MovieGenre=Action\&Triller}> (i.e., red dotted rectangle). This is very reasonable since young men are very likely to prefer action\&triller movies. 


We are also interested in what the correlations between different feature fields in the data are. Therefore, we measure the correlations between the feature fields according to their average attention score in the entire data. The correlations between different fields are summarized into Figure~\ref{fig::case} (b). We can see that <\textit{Gender, Genre}>, <\textit{Age, Genre}>, <\textit{RequestTime, ReleaseTime}> and <\textit{Gender, Age, Genre}> (i.e., solid green region) are strongly correlated, which are the explainable rules for recommendation in this domain. 


\vspace{-2pt}
\subsection{Integrating Implicit Interactions (RQ4)}\label{sec::joint}
Feed-forward neural networks are capable of modeling implicit feature interactions and have been widely integrated into existing CTR prediction methods~\cite{cheng2016wide,guo2017deepfm,lian2018xdeepfm}. To investigate whether integrating implicit feature interactions further improves the performance, we combine AutoInt with a two-layer feed-forward neural network by joint training. 
We name the joint model \textit{AutoInt+} and compare it with the following algorithms:
\begin{itemize}[leftmargin=*]
\item Wide\&Deep~\cite{cheng2016wide}. Wide\&Deep integrates the outputs of logistic regression and feed-forward neural networks.
\item DeepFM~\cite{guo2017deepfm}. DeepFM combines trainditional second-order factorization machines and feed-forward neural network, with a shared embedding layer.
\item Deep\&Cross~\cite{wang2017deep}. Deep\&Cross is the extension of CrossNet by integrating feed-forward neural networks.
\item xDeepFM~\cite{lian2018xdeepfm}. xDeepFM is the extension of CIN by integrating feed-forward neural networks.
\end{itemize}

Table~\ref{tab::ensemble} presents the averaged results (over 10 runs) of joint-training models. We have the following observations: 1) The performance of our method improves by joint training with feed-forward neural networks on all datasets. This indicates that integrating implicit feature interactions indeed boosts the predictive ability of our proposed model. However, as can be seen from last two columns, the magnitude of performance improvement is fairly small compared to other models, showing that our individual model AutoInt is quite powerful. 2) After integrating implicit feature interactions, AutoInt+ outperforms all competitive methods, and achieves new state-of-the-art performances on used CTR prediction data sets. 

%% file: conclusion.tex
\section{Conclusion and future work}
In this work, we propose a novel CTR prediction model based on self-attention mechanism, which can automatically learn high-order feature interactions in an explicit fashion. The key to our method is the newly-introduced interacting layer, which allows each feature to interact with the others and to determine the relevance through learning. Experimental results on four real-world data sets demonstrate the effectiveness and efficiency of our proposed model. Besides, we provide good model explainability via visualizing the learned combinatorial features. When integrating with implicit feature interactions captured by feed-forward neural networks, we achieve better offline AUC and \textit{Logloss} scores compared to the previous state-of-the-art methods. 

For future work , we are interested in incorporating contextual information into our method and improving its performance for online recommender systems. Besides, we also plan to extend AutoInt for general machine learning tasks, such as regression, classification and ranking. 

%% file: ack.tex
\section{Acknowledgement}
The authors would like to thank all the anonymous reviewers for their insightful comments. We thank Xiao Xiao and Jianbo Dong for the discussion on recommendation mechanism in China University MOOC platform.
We also thank Meng Qu for reviewing the initial version of this paper. 
Weiping Song and Ming Zhang are supported by National Key Research and Development Program of China with Grant No. SQ2018AAA010010, Beijing Municipal Commission of Science and Technology under Grant No. Z181100008918005 as well as the National Natural Science Foundation of China (NSFC Grant Nos.61772039 and 91646202). Weiping Song is also supported by Chinese Scholarship Council. Jian Tang is supported by the Natural Sciences and Engineering Research Council of Canada, as well as the Canada CIFAR AI Chair Program.

%% file: main.bbl

\begin{thebibliography}{44}


\ifx \showCODEN    \undefined \def \showCODEN     #1{\unskip}     \fi
\ifx \showDOI      \undefined \def \showDOI       #1{#1}\fi
\ifx \showISBNx    \undefined \def \showISBNx     #1{\unskip}     \fi
\ifx \showISBNxiii \undefined \def \showISBNxiii  #1{\unskip}     \fi
\ifx \showISSN     \undefined \def \showISSN      #1{\unskip}     \fi
\ifx \showLCCN     \undefined \def \showLCCN      #1{\unskip}     \fi
\ifx \shownote     \undefined \def \shownote      #1{#1}          \fi
\ifx \showarticletitle \undefined \def \showarticletitle #1{#1}   \fi
\ifx \showURL      \undefined \def \showURL       {\relax}        \fi
\providecommand\bibfield[2]{#2}
\providecommand\bibinfo[2]{#2}
\providecommand\natexlab[1]{#1}
\providecommand\showeprint[2][]{arXiv:#2}

\bibitem[\protect\citeauthoryear{Abadi, Barham, Chen, Chen, Davis, Dean, Devin,
  Ghemawat, Irving, et~al\mbox{.}}{Abadi et~al\mbox{.}}{2016}]%
        {abadi2016tensorflow}
\bibfield{author}{\bibinfo{person}{Mart{\'\i}n Abadi}, \bibinfo{person}{Paul
  Barham}, \bibinfo{person}{Jianmin Chen}, \bibinfo{person}{Zhifeng Chen},
  \bibinfo{person}{Andy Davis}, \bibinfo{person}{Jeffrey Dean},
  \bibinfo{person}{Matthieu Devin}, \bibinfo{person}{Sanjay Ghemawat},
  \bibinfo{person}{Geoffrey Irving}, {et~al\mbox{.}}}
  \bibinfo{year}{2016}\natexlab{}.
\newblock \showarticletitle{TensorFlow: A System for Large-Scale Machine
  Learning.}. In \bibinfo{booktitle}{\emph{OSDI}}, Vol.~\bibinfo{volume}{16}.
  \bibinfo{pages}{265--283}.
\newblock


\bibitem[\protect\citeauthoryear{Bahdanau, Cho, and Bengio}{Bahdanau
  et~al\mbox{.}}{2015}]%
        {bahdanau2014neural}
\bibfield{author}{\bibinfo{person}{Dzmitry Bahdanau},
  \bibinfo{person}{Kyunghyun Cho}, {and} \bibinfo{person}{Yoshua Bengio}.}
  \bibinfo{year}{2015}\natexlab{}.
\newblock \showarticletitle{Neural machine translation by jointly learning to
  align and translate}. In \bibinfo{booktitle}{\emph{International Conference
  on Learning Representations}}.
\newblock


\bibitem[\protect\citeauthoryear{Bengio, Courville, and Vincent}{Bengio
  et~al\mbox{.}}{2013}]%
        {bengio2013representation}
\bibfield{author}{\bibinfo{person}{Yoshua Bengio}, \bibinfo{person}{Aaron
  Courville}, {and} \bibinfo{person}{Pascal Vincent}.}
  \bibinfo{year}{2013}\natexlab{}.
\newblock \showarticletitle{Representation learning: A review and new
  perspectives}.
\newblock \bibinfo{journal}{\emph{IEEE transactions on pattern analysis and
  machine intelligence}} \bibinfo{volume}{35}, \bibinfo{number}{8}
  (\bibinfo{year}{2013}), \bibinfo{pages}{1798--1828}.
\newblock


\bibitem[\protect\citeauthoryear{Beutel, Covington, Jain, Xu, Li, Gatto, and
  Chi}{Beutel et~al\mbox{.}}{2018}]%
        {beutel2018latent}
\bibfield{author}{\bibinfo{person}{Alex Beutel}, \bibinfo{person}{Paul
  Covington}, \bibinfo{person}{Sagar Jain}, \bibinfo{person}{Can Xu},
  \bibinfo{person}{Jia Li}, \bibinfo{person}{Vince Gatto}, {and}
  \bibinfo{person}{Ed~H Chi}.} \bibinfo{year}{2018}\natexlab{}.
\newblock \showarticletitle{Latent Cross: Making Use of Context in Recurrent
  Recommender Systems}. In \bibinfo{booktitle}{\emph{Proceedings of the
  Eleventh ACM International Conference on Web Search and Data Mining}}. ACM,
  \bibinfo{pages}{46--54}.
\newblock


\bibitem[\protect\citeauthoryear{Blondel, Fujino, Ueda, and Ishihata}{Blondel
  et~al\mbox{.}}{2016a}]%
        {blondel2016higher}
\bibfield{author}{\bibinfo{person}{Mathieu Blondel}, \bibinfo{person}{Akinori
  Fujino}, \bibinfo{person}{Naonori Ueda}, {and} \bibinfo{person}{Masakazu
  Ishihata}.} \bibinfo{year}{2016}\natexlab{a}.
\newblock \showarticletitle{Higher-order factorization machines}. In
  \bibinfo{booktitle}{\emph{Advances in Neural Information Processing
  Systems}}. \bibinfo{pages}{3351--3359}.
\newblock


\bibitem[\protect\citeauthoryear{Blondel, Ishihata, Fujino, and Ueda}{Blondel
  et~al\mbox{.}}{2016b}]%
        {blondel2016polynomial}
\bibfield{author}{\bibinfo{person}{Mathieu Blondel}, \bibinfo{person}{Masakazu
  Ishihata}, \bibinfo{person}{Akinori Fujino}, {and} \bibinfo{person}{Naonori
  Ueda}.} \bibinfo{year}{2016}\natexlab{b}.
\newblock \showarticletitle{Polynomial Networks and Factorization Machines: New
  Insights and Efficient Training Algorithms}. In
  \bibinfo{booktitle}{\emph{International Conference on Machine Learning}}.
  \bibinfo{pages}{850--858}.
\newblock


\bibitem[\protect\citeauthoryear{Cheng, Xia, Zhang, King, and Lyu}{Cheng
  et~al\mbox{.}}{2014}]%
        {cheng2014gradient}
\bibfield{author}{\bibinfo{person}{Chen Cheng}, \bibinfo{person}{Fen Xia},
  \bibinfo{person}{Tong Zhang}, \bibinfo{person}{Irwin King}, {and}
  \bibinfo{person}{Michael~R Lyu}.} \bibinfo{year}{2014}\natexlab{}.
\newblock \showarticletitle{Gradient boosting factorization machines}. In
  \bibinfo{booktitle}{\emph{Proceedings of the 8th ACM Conference on
  Recommender systems}}. ACM, \bibinfo{pages}{265--272}.
\newblock


\bibitem[\protect\citeauthoryear{Cheng, Koc, Harmsen, Shaked, Chandra, Aradhye,
  Anderson, Corrado, Chai, Ispir, et~al\mbox{.}}{Cheng et~al\mbox{.}}{2016}]%
        {cheng2016wide}
\bibfield{author}{\bibinfo{person}{Heng-Tze Cheng}, \bibinfo{person}{Levent
  Koc}, \bibinfo{person}{Jeremiah Harmsen}, \bibinfo{person}{Tal Shaked},
  \bibinfo{person}{Tushar Chandra}, \bibinfo{person}{Hrishi Aradhye},
  \bibinfo{person}{Glen Anderson}, \bibinfo{person}{Greg Corrado},
  \bibinfo{person}{Wei Chai}, \bibinfo{person}{Mustafa Ispir}, {et~al\mbox{.}}}
  \bibinfo{year}{2016}\natexlab{}.
\newblock \showarticletitle{Wide \& deep learning for recommender systems}. In
  \bibinfo{booktitle}{\emph{Proceedings of the 1st Workshop on Deep Learning
  for Recommender Systems}}. ACM, \bibinfo{pages}{7--10}.
\newblock


\bibitem[\protect\citeauthoryear{Covington, Adams, and Sargin}{Covington
  et~al\mbox{.}}{2016}]%
        {covington2016deep}
\bibfield{author}{\bibinfo{person}{Paul Covington}, \bibinfo{person}{Jay
  Adams}, {and} \bibinfo{person}{Emre Sargin}.}
  \bibinfo{year}{2016}\natexlab{}.
\newblock \showarticletitle{Deep neural networks for youtube recommendations}.
  In \bibinfo{booktitle}{\emph{Proceedings of the 10th ACM Conference on
  Recommender Systems}}. ACM, \bibinfo{pages}{191--198}.
\newblock


\bibitem[\protect\citeauthoryear{Graepel, Candela, Borchert, and
  Herbrich}{Graepel et~al\mbox{.}}{2010}]%
        {graepel2010web}
\bibfield{author}{\bibinfo{person}{Thore Graepel}, \bibinfo{person}{Joaquin
  Qui\~{n}onero Candela}, \bibinfo{person}{Thomas Borchert}, {and}
  \bibinfo{person}{Ralf Herbrich}.} \bibinfo{year}{2010}\natexlab{}.
\newblock \showarticletitle{Web-scale Bayesian Click-through Rate Prediction
  for Sponsored Search Advertising in Microsoft's Bing Search Engine}. In
  \bibinfo{booktitle}{\emph{Proceedings of the 27th International Conference on
  International Conference on Machine Learning}}. \bibinfo{pages}{13--20}.
\newblock


\bibitem[\protect\citeauthoryear{Guo, Tang, Ye, Li, and He}{Guo
  et~al\mbox{.}}{2017}]%
        {guo2017deepfm}
\bibfield{author}{\bibinfo{person}{Huifeng Guo}, \bibinfo{person}{Ruiming
  Tang}, \bibinfo{person}{Yunming Ye}, \bibinfo{person}{Zhenguo Li}, {and}
  \bibinfo{person}{Xiuqiang He}.} \bibinfo{year}{2017}\natexlab{}.
\newblock \showarticletitle{DeepFM: A Factorization-machine Based Neural
  Network for CTR Prediction}. In \bibinfo{booktitle}{\emph{Proceedings of the
  26th International Joint Conference on Artificial Intelligence}}.
  \bibinfo{publisher}{AAAI Press}, \bibinfo{pages}{1725--1731}.
\newblock


\bibitem[\protect\citeauthoryear{He, Zhang, Ren, and Sun}{He
  et~al\mbox{.}}{2016}]%
        {he2016deep}
\bibfield{author}{\bibinfo{person}{Kaiming He}, \bibinfo{person}{Xiangyu
  Zhang}, \bibinfo{person}{Shaoqing Ren}, {and} \bibinfo{person}{Jian Sun}.}
  \bibinfo{year}{2016}\natexlab{}.
\newblock \showarticletitle{Deep residual learning for image recognition}. In
  \bibinfo{booktitle}{\emph{Proceedings of the IEEE conference on computer
  vision and pattern recognition}}. \bibinfo{pages}{770--778}.
\newblock


\bibitem[\protect\citeauthoryear{He and Chua}{He and Chua}{2017}]%
        {he2017neural}
\bibfield{author}{\bibinfo{person}{Xiangnan He} {and} \bibinfo{person}{Tat-Seng
  Chua}.} \bibinfo{year}{2017}\natexlab{}.
\newblock \showarticletitle{Neural factorization machines for sparse predictive
  analytics}. In \bibinfo{booktitle}{\emph{Proceedings of the 40th
  International ACM SIGIR conference on Research and Development in Information
  Retrieval}}. ACM, \bibinfo{pages}{355--364}.
\newblock


\bibitem[\protect\citeauthoryear{He, He, Song, Liu, Jiang, and Chua}{He
  et~al\mbox{.}}{2018}]%
        {he2018nais}
\bibfield{author}{\bibinfo{person}{Xiangnan He}, \bibinfo{person}{Zhankui He},
  \bibinfo{person}{Jingkuan Song}, \bibinfo{person}{Zhenguang Liu},
  \bibinfo{person}{Yu-Gang Jiang}, {and} \bibinfo{person}{Tat-Seng Chua}.}
  \bibinfo{year}{2018}\natexlab{}.
\newblock \showarticletitle{NAIS: Neural attentive item similarity model for
  recommendation}.
\newblock \bibinfo{journal}{\emph{IEEE Transactions on Knowledge and Data
  Engineering}} \bibinfo{volume}{30}, \bibinfo{number}{12}
  (\bibinfo{year}{2018}), \bibinfo{pages}{2354--2366}.
\newblock


\bibitem[\protect\citeauthoryear{He, Pan, Jin, Xu, Liu, Xu, Shi, Atallah,
  Herbrich, Bowers, et~al\mbox{.}}{He et~al\mbox{.}}{2014}]%
        {he2014practical}
\bibfield{author}{\bibinfo{person}{Xinran He}, \bibinfo{person}{Junfeng Pan},
  \bibinfo{person}{Ou Jin}, \bibinfo{person}{Tianbing Xu}, \bibinfo{person}{Bo
  Liu}, \bibinfo{person}{Tao Xu}, \bibinfo{person}{Yanxin Shi},
  \bibinfo{person}{Antoine Atallah}, \bibinfo{person}{Ralf Herbrich},
  \bibinfo{person}{Stuart Bowers}, {et~al\mbox{.}}}
  \bibinfo{year}{2014}\natexlab{}.
\newblock \showarticletitle{Practical lessons from predicting clicks on ads at
  facebook}. In \bibinfo{booktitle}{\emph{Proceedings of the Eighth
  International Workshop on Data Mining for Online Advertising}}. ACM,
  \bibinfo{pages}{1--9}.
\newblock


\bibitem[\protect\citeauthoryear{Juan, Zhuang, Chin, and Lin}{Juan
  et~al\mbox{.}}{2016}]%
        {juan2016field}
\bibfield{author}{\bibinfo{person}{Yuchin Juan}, \bibinfo{person}{Yong Zhuang},
  \bibinfo{person}{Wei-Sheng Chin}, {and} \bibinfo{person}{Chih-Jen Lin}.}
  \bibinfo{year}{2016}\natexlab{}.
\newblock \showarticletitle{Field-aware factorization machines for CTR
  prediction}. In \bibinfo{booktitle}{\emph{Proceedings of the 10th ACM
  Conference on Recommender Systems}}. ACM, \bibinfo{pages}{43--50}.
\newblock


\bibitem[\protect\citeauthoryear{Kingma and Ba}{Kingma and Ba}{2015}]%
        {kingma2014adam}
\bibfield{author}{\bibinfo{person}{Diederick~P Kingma} {and}
  \bibinfo{person}{Jimmy Ba}.} \bibinfo{year}{2015}\natexlab{}.
\newblock \showarticletitle{Adam: A method for stochastic optimization}. In
  \bibinfo{booktitle}{\emph{International Conference on Learning
  Representations}}.
\newblock


\bibitem[\protect\citeauthoryear{Lee, Grosse, Ranganath, and Ng}{Lee
  et~al\mbox{.}}{2011}]%
        {lee2011unsupervised}
\bibfield{author}{\bibinfo{person}{Honglak Lee}, \bibinfo{person}{Roger
  Grosse}, \bibinfo{person}{Rajesh Ranganath}, {and} \bibinfo{person}{Andrew~Y
  Ng}.} \bibinfo{year}{2011}\natexlab{}.
\newblock \showarticletitle{Unsupervised learning of hierarchical
  representations with convolutional deep belief networks}.
\newblock \bibinfo{journal}{\emph{Commun. ACM}} \bibinfo{volume}{54},
  \bibinfo{number}{10} (\bibinfo{year}{2011}), \bibinfo{pages}{95--103}.
\newblock


\bibitem[\protect\citeauthoryear{Lian, Zhou, Zhang, Chen, Xie, and Sun}{Lian
  et~al\mbox{.}}{2018}]%
        {lian2018xdeepfm}
\bibfield{author}{\bibinfo{person}{Jianxun Lian}, \bibinfo{person}{Xiaohuan
  Zhou}, \bibinfo{person}{Fuzheng Zhang}, \bibinfo{person}{Zhongxia Chen},
  \bibinfo{person}{Xing Xie}, {and} \bibinfo{person}{Guangzhong Sun}.}
  \bibinfo{year}{2018}\natexlab{}.
\newblock \showarticletitle{xDeepFM: Combining Explicit and Implicit Feature
  Interactions for Recommender Systems}. In
  \bibinfo{booktitle}{\emph{Proceedings of the 24th ACM SIGKDD International
  Conference on Knowledge Discovery and Data Mining}}.
  \bibinfo{publisher}{ACM}, \bibinfo{pages}{1754--1763}.
\newblock


\bibitem[\protect\citeauthoryear{Lin, Feng, Santos, Yu, Xiang, Zhou, and
  Bengio}{Lin et~al\mbox{.}}{2017}]%
        {lin2017structured}
\bibfield{author}{\bibinfo{person}{Zhouhan Lin}, \bibinfo{person}{Minwei Feng},
  \bibinfo{person}{Cicero Nogueira~dos Santos}, \bibinfo{person}{Mo Yu},
  \bibinfo{person}{Bing Xiang}, \bibinfo{person}{Bowen Zhou}, {and}
  \bibinfo{person}{Yoshua Bengio}.} \bibinfo{year}{2017}\natexlab{}.
\newblock \showarticletitle{A structured self-attentive sentence embedding}. In
  \bibinfo{booktitle}{\emph{International Conference on Learning
  Representations}}.
\newblock


\bibitem[\protect\citeauthoryear{McMahan, Holt, Sculley, Young, Ebner, Grady,
  Nie, Phillips, et~al\mbox{.}}{McMahan et~al\mbox{.}}{2013}]%
        {mcmahan2013ad}
\bibfield{author}{\bibinfo{person}{H.~Brendan McMahan}, \bibinfo{person}{Gary
  Holt}, \bibinfo{person}{D. Sculley}, \bibinfo{person}{Michael Young},
  \bibinfo{person}{Dietmar Ebner}, \bibinfo{person}{Julian Grady},
  \bibinfo{person}{Lan Nie}, \bibinfo{person}{Todd Phillips}, {et~al\mbox{.}}}
  \bibinfo{year}{2013}\natexlab{}.
\newblock \showarticletitle{Ad Click Prediction: A View from the Trenches}. In
  \bibinfo{booktitle}{\emph{Proceedings of the 19th ACM SIGKDD International
  Conference on Knowledge Discovery and Data Mining}}.
  \bibinfo{publisher}{ACM}, \bibinfo{pages}{1222--1230}.
\newblock


\bibitem[\protect\citeauthoryear{Miller, Fisch, Dodge, Karimi, Bordes, and
  Weston}{Miller et~al\mbox{.}}{2016}]%
        {miller2016key}
\bibfield{author}{\bibinfo{person}{Alexander Miller}, \bibinfo{person}{Adam
  Fisch}, \bibinfo{person}{Jesse Dodge}, \bibinfo{person}{Amir-Hossein Karimi},
  \bibinfo{person}{Antoine Bordes}, {and} \bibinfo{person}{Jason Weston}.}
  \bibinfo{year}{2016}\natexlab{}.
\newblock \showarticletitle{Key-Value Memory Networks for Directly Reading
  Documents}. In \bibinfo{booktitle}{\emph{Proceedings of the 2016 Conference
  on Empirical Methods in Natural Language Processing}}.
  \bibinfo{publisher}{Association for Computational Linguistics},
  \bibinfo{pages}{1400--1409}.
\newblock


\bibitem[\protect\citeauthoryear{Novikov, Trofimov, and Oseledets}{Novikov
  et~al\mbox{.}}{2016}]%
        {novikov2016exponential}
\bibfield{author}{\bibinfo{person}{Alexander Novikov}, \bibinfo{person}{Mikhail
  Trofimov}, {and} \bibinfo{person}{Ivan Oseledets}.}
  \bibinfo{year}{2016}\natexlab{}.
\newblock \showarticletitle{Exponential machines}.
\newblock \bibinfo{journal}{\emph{arXiv preprint arXiv:1605.03795}}
  (\bibinfo{year}{2016}).
\newblock


\bibitem[\protect\citeauthoryear{Oentaryo, Lim, Low, Lo, and Finegold}{Oentaryo
  et~al\mbox{.}}{2014}]%
        {oentaryo2014predicting}
\bibfield{author}{\bibinfo{person}{Richard~J Oentaryo},
  \bibinfo{person}{Ee-Peng Lim}, \bibinfo{person}{Jia-Wei Low},
  \bibinfo{person}{David Lo}, {and} \bibinfo{person}{Michael Finegold}.}
  \bibinfo{year}{2014}\natexlab{}.
\newblock \showarticletitle{Predicting response in mobile advertising with
  hierarchical importance-aware factorization machine}. In
  \bibinfo{booktitle}{\emph{Proceedings of the 7th ACM international conference
  on Web search and data mining}}. ACM, \bibinfo{pages}{123--132}.
\newblock


\bibitem[\protect\citeauthoryear{Qu, Cai, Ren, Zhang, Yu, Wen, and Wang}{Qu
  et~al\mbox{.}}{2016}]%
        {qu2016product}
\bibfield{author}{\bibinfo{person}{Yanru Qu}, \bibinfo{person}{Han Cai},
  \bibinfo{person}{Kan Ren}, \bibinfo{person}{Weinan Zhang},
  \bibinfo{person}{Yong Yu}, \bibinfo{person}{Ying Wen}, {and}
  \bibinfo{person}{Jun Wang}.} \bibinfo{year}{2016}\natexlab{}.
\newblock \showarticletitle{Product-based neural networks for user response
  prediction}. In \bibinfo{booktitle}{\emph{Data Mining (ICDM), 2016 IEEE 16th
  International Conference on}}. IEEE, \bibinfo{pages}{1149--1154}.
\newblock


\bibitem[\protect\citeauthoryear{Rendle}{Rendle}{2010}]%
        {rendle2010factorization}
\bibfield{author}{\bibinfo{person}{Steffen Rendle}.}
  \bibinfo{year}{2010}\natexlab{}.
\newblock \showarticletitle{Factorization machines}. In
  \bibinfo{booktitle}{\emph{Data Mining (ICDM), 2010 IEEE 10th International
  Conference on}}. IEEE, \bibinfo{pages}{995--1000}.
\newblock


\bibitem[\protect\citeauthoryear{Rendle, Freudenthaler, and
  Schmidt-Thieme}{Rendle et~al\mbox{.}}{2010}]%
        {rendle2010factorizing}
\bibfield{author}{\bibinfo{person}{Steffen Rendle}, \bibinfo{person}{Christoph
  Freudenthaler}, {and} \bibinfo{person}{Lars Schmidt-Thieme}.}
  \bibinfo{year}{2010}\natexlab{}.
\newblock \showarticletitle{Factorizing personalized markov chains for
  next-basket recommendation}. In \bibinfo{booktitle}{\emph{Proceedings of the
  19th international conference on World wide web}}. ACM,
  \bibinfo{pages}{811--820}.
\newblock


\bibitem[\protect\citeauthoryear{Rendle, Gantner, Freudenthaler, and
  Schmidt-Thieme}{Rendle et~al\mbox{.}}{2011}]%
        {rendle2011fast}
\bibfield{author}{\bibinfo{person}{Steffen Rendle}, \bibinfo{person}{Zeno
  Gantner}, \bibinfo{person}{Christoph Freudenthaler}, {and}
  \bibinfo{person}{Lars Schmidt-Thieme}.} \bibinfo{year}{2011}\natexlab{}.
\newblock \showarticletitle{Fast context-aware recommendations with
  factorization machines}. In \bibinfo{booktitle}{\emph{Proceedings of the 34th
  international ACM SIGIR conference on Research and development in Information
  Retrieval}}. ACM, \bibinfo{pages}{635--644}.
\newblock


\bibitem[\protect\citeauthoryear{Richardson, Dominowska, and Ragno}{Richardson
  et~al\mbox{.}}{2007}]%
        {richardson2007predicting}
\bibfield{author}{\bibinfo{person}{Matthew Richardson}, \bibinfo{person}{Ewa
  Dominowska}, {and} \bibinfo{person}{Robert Ragno}.}
  \bibinfo{year}{2007}\natexlab{}.
\newblock \showarticletitle{Predicting clicks: estimating the click-through
  rate for new ads}. In \bibinfo{booktitle}{\emph{Proceedings of the 16th
  international conference on World Wide Web}}. ACM, \bibinfo{pages}{521--530}.
\newblock


\bibitem[\protect\citeauthoryear{Rush, Chopra, and Weston}{Rush
  et~al\mbox{.}}{2015}]%
        {rush2015neural}
\bibfield{author}{\bibinfo{person}{Alexander~M. Rush}, \bibinfo{person}{Sumit
  Chopra}, {and} \bibinfo{person}{Jason Weston}.}
  \bibinfo{year}{2015}\natexlab{}.
\newblock \showarticletitle{A Neural Attention Model for Abstractive Sentence
  Summarization}. In \bibinfo{booktitle}{\emph{Proceedings of the 2015
  Conference on Empirical Methods in Natural Language Processing}}.
  \bibinfo{publisher}{Association for Computational Linguistics},
  \bibinfo{pages}{379--389}.
\newblock


\bibitem[\protect\citeauthoryear{Shan, Lin, Sun, and Wang}{Shan
  et~al\mbox{.}}{2016b}]%
        {shan2016predicting}
\bibfield{author}{\bibinfo{person}{Lili Shan}, \bibinfo{person}{Lei Lin},
  \bibinfo{person}{Chengjie Sun}, {and} \bibinfo{person}{Xiaolong Wang}.}
  \bibinfo{year}{2016}\natexlab{b}.
\newblock \showarticletitle{Predicting ad click-through rates via feature-based
  fully coupled interaction tensor factorization}.
\newblock \bibinfo{journal}{\emph{Electronic Commerce Research and
  Applications}}  \bibinfo{volume}{16} (\bibinfo{year}{2016}),
  \bibinfo{pages}{30--42}.
\newblock


\bibitem[\protect\citeauthoryear{Shan, Hoens, Jiao, Wang, Yu, and Mao}{Shan
  et~al\mbox{.}}{2016a}]%
        {shan2016deep}
\bibfield{author}{\bibinfo{person}{Ying Shan}, \bibinfo{person}{T~Ryan Hoens},
  \bibinfo{person}{Jian Jiao}, \bibinfo{person}{Haijing Wang},
  \bibinfo{person}{Dong Yu}, {and} \bibinfo{person}{JC Mao}.}
  \bibinfo{year}{2016}\natexlab{a}.
\newblock \showarticletitle{Deep crossing: Web-scale modeling without manually
  crafted combinatorial features}. In \bibinfo{booktitle}{\emph{Proceedings of
  the 22nd ACM SIGKDD International Conference on Knowledge Discovery and Data
  Mining}}. ACM, \bibinfo{pages}{255--262}.
\newblock


\bibitem[\protect\citeauthoryear{Song, Xiao, Wang, Charlin, Zhang, and
  Tang}{Song et~al\mbox{.}}{2019}]%
        {song2019session}
\bibfield{author}{\bibinfo{person}{Weiping Song}, \bibinfo{person}{Zhiping
  Xiao}, \bibinfo{person}{Yifan Wang}, \bibinfo{person}{Laurent Charlin},
  \bibinfo{person}{Ming Zhang}, {and} \bibinfo{person}{Jian Tang}.}
  \bibinfo{year}{2019}\natexlab{}.
\newblock \showarticletitle{Session-based Social Recommendation via Dynamic
  Graph Attention Networks}. In \bibinfo{booktitle}{\emph{Proceedings of the
  Twelfth ACM International Conference on Web Search and Data Mining}}. ACM,
  \bibinfo{pages}{555--563}.
\newblock


\bibitem[\protect\citeauthoryear{Srivastava, Hinton, Krizhevsky, Sutskever, and
  Salakhutdinov}{Srivastava et~al\mbox{.}}{2014}]%
        {srivastava2014dropout}
\bibfield{author}{\bibinfo{person}{Nitish Srivastava},
  \bibinfo{person}{Geoffrey Hinton}, \bibinfo{person}{Alex Krizhevsky},
  \bibinfo{person}{Ilya Sutskever}, {and} \bibinfo{person}{Ruslan
  Salakhutdinov}.} \bibinfo{year}{2014}\natexlab{}.
\newblock \showarticletitle{Dropout: A simple way to prevent neural networks
  from overfitting}.
\newblock \bibinfo{journal}{\emph{The Journal of Machine Learning Research}}
  \bibinfo{volume}{15}, \bibinfo{number}{1} (\bibinfo{year}{2014}),
  \bibinfo{pages}{1929--1958}.
\newblock


\bibitem[\protect\citeauthoryear{Sukhbaatar, Weston, Fergus,
  et~al\mbox{.}}{Sukhbaatar et~al\mbox{.}}{2015}]%
        {sukhbaatar2015end}
\bibfield{author}{\bibinfo{person}{Sainbayar Sukhbaatar},
  \bibinfo{person}{Jason Weston}, \bibinfo{person}{Rob Fergus},
  {et~al\mbox{.}}} \bibinfo{year}{2015}\natexlab{}.
\newblock \showarticletitle{End-to-end memory networks}. In
  \bibinfo{booktitle}{\emph{Advances in neural information processing
  systems}}. \bibinfo{pages}{2440--2448}.
\newblock


\bibitem[\protect\citeauthoryear{Vaswani, Shazeer, Parmar, Uszkoreit, Jones,
  Gomez, Kaiser, and Polosukhin}{Vaswani et~al\mbox{.}}{2017}]%
        {vaswani2017attention}
\bibfield{author}{\bibinfo{person}{Ashish Vaswani}, \bibinfo{person}{Noam
  Shazeer}, \bibinfo{person}{Niki Parmar}, \bibinfo{person}{Jakob Uszkoreit},
  \bibinfo{person}{Llion Jones}, \bibinfo{person}{Aidan~N Gomez},
  \bibinfo{person}{{\L}ukasz Kaiser}, {and} \bibinfo{person}{Illia
  Polosukhin}.} \bibinfo{year}{2017}\natexlab{}.
\newblock \showarticletitle{Attention is all you need}. In
  \bibinfo{booktitle}{\emph{Advances in Neural Information Processing
  Systems}}. \bibinfo{pages}{6000--6010}.
\newblock


\bibitem[\protect\citeauthoryear{Velickovic, Cucurull, Casanova, Romero, Lio,
  and Bengio}{Velickovic et~al\mbox{.}}{2018}]%
        {velickovic2017graph}
\bibfield{author}{\bibinfo{person}{Petar Velickovic}, \bibinfo{person}{Guillem
  Cucurull}, \bibinfo{person}{Arantxa Casanova}, \bibinfo{person}{Adriana
  Romero}, \bibinfo{person}{Pietro Lio}, {and} \bibinfo{person}{Yoshua
  Bengio}.} \bibinfo{year}{2018}\natexlab{}.
\newblock \showarticletitle{Graph Attention Networks}. In
  \bibinfo{booktitle}{\emph{International Conference on Learning
  Representations}}.
\newblock


\bibitem[\protect\citeauthoryear{Wang, Fu, Fu, and Wang}{Wang
  et~al\mbox{.}}{2017}]%
        {wang2017deep}
\bibfield{author}{\bibinfo{person}{Ruoxi Wang}, \bibinfo{person}{Bin Fu},
  \bibinfo{person}{Gang Fu}, {and} \bibinfo{person}{Mingliang Wang}.}
  \bibinfo{year}{2017}\natexlab{}.
\newblock \showarticletitle{Deep \& Cross Network for Ad Click Predictions}. In
  \bibinfo{booktitle}{\emph{Proceedings of the ADKDD'17}}.
  \bibinfo{publisher}{ACM}, \bibinfo{pages}{12:1--12:7}.
\newblock


\bibitem[\protect\citeauthoryear{Wang, He, Feng, Nie, and Chua}{Wang
  et~al\mbox{.}}{2018}]%
        {wang2018tem}
\bibfield{author}{\bibinfo{person}{Xiang Wang}, \bibinfo{person}{Xiangnan He},
  \bibinfo{person}{Fuli Feng}, \bibinfo{person}{Liqiang Nie}, {and}
  \bibinfo{person}{Tat-Seng Chua}.} \bibinfo{year}{2018}\natexlab{}.
\newblock \showarticletitle{TEM: Tree-enhanced Embedding Model for Explainable
  Recommendation}. In \bibinfo{booktitle}{\emph{Proceedings of the 2018 World
  Wide Web Conference on World Wide Web}}. International World Wide Web
  Conferences Steering Committee, \bibinfo{pages}{1543--1552}.
\newblock


\bibitem[\protect\citeauthoryear{Xiao, Ye, He, Zhang, Wu, and Chua}{Xiao
  et~al\mbox{.}}{2017}]%
        {xiao2017attentional}
\bibfield{author}{\bibinfo{person}{Jun Xiao}, \bibinfo{person}{Hao Ye},
  \bibinfo{person}{Xiangnan He}, \bibinfo{person}{Hanwang Zhang},
  \bibinfo{person}{Fei Wu}, {and} \bibinfo{person}{Tat-Seng Chua}.}
  \bibinfo{year}{2017}\natexlab{}.
\newblock \showarticletitle{Attentional factorization machines: learning the
  weight of feature interactions via attention networks}. In
  \bibinfo{booktitle}{\emph{Proceedings of the 26th International Joint
  Conference on Artificial Intelligence}}. AAAI Press,
  \bibinfo{pages}{3119--3125}.
\newblock


\bibitem[\protect\citeauthoryear{Zhang, Du, and Wang}{Zhang
  et~al\mbox{.}}{2016}]%
        {zhang2016deep}
\bibfield{author}{\bibinfo{person}{Weinan Zhang}, \bibinfo{person}{Tianming
  Du}, {and} \bibinfo{person}{Jun Wang}.} \bibinfo{year}{2016}\natexlab{}.
\newblock \showarticletitle{Deep learning over multi-field categorical data}.
  In \bibinfo{booktitle}{\emph{European conference on information retrieval}}.
  Springer, \bibinfo{pages}{45--57}.
\newblock


\bibitem[\protect\citeauthoryear{Zhao, Shi, and Hong}{Zhao
  et~al\mbox{.}}{2017}]%
        {zhao2017gb}
\bibfield{author}{\bibinfo{person}{Qian Zhao}, \bibinfo{person}{Yue Shi}, {and}
  \bibinfo{person}{Liangjie Hong}.} \bibinfo{year}{2017}\natexlab{}.
\newblock \showarticletitle{GB-CENT: Gradient Boosted Categorical Embedding and
  Numerical Trees}. In \bibinfo{booktitle}{\emph{Proceedings of the 26th
  International Conference on World Wide Web}}. International World Wide Web
  Conferences Steering Committee, \bibinfo{pages}{1311--1319}.
\newblock


\bibitem[\protect\citeauthoryear{Zhou, Zhu, Song, Fan, Zhu, Ma, Yan, Jin, Li,
  and Gai}{Zhou et~al\mbox{.}}{2018}]%
        {zhou2017deep}
\bibfield{author}{\bibinfo{person}{Guorui Zhou}, \bibinfo{person}{Xiaoqiang
  Zhu}, \bibinfo{person}{Chenru Song}, \bibinfo{person}{Ying Fan},
  \bibinfo{person}{Han Zhu}, \bibinfo{person}{Xiao Ma},
  \bibinfo{person}{Yanghui Yan}, \bibinfo{person}{Junqi Jin},
  \bibinfo{person}{Han Li}, {and} \bibinfo{person}{Kun Gai}.}
  \bibinfo{year}{2018}\natexlab{}.
\newblock \showarticletitle{Deep Interest Network for Click-Through Rate
  Prediction}. In \bibinfo{booktitle}{\emph{Proceedings of the 24th ACM SIGKDD
  International Conference on Knowledge Discovery and Data Mining}}.
  \bibinfo{publisher}{ACM}, \bibinfo{pages}{1059--1068}.
\newblock


\bibitem[\protect\citeauthoryear{Zhu, Shan, Mao, Yu, Rahmanian, and Zhang}{Zhu
  et~al\mbox{.}}{2017}]%
        {zhu2017deep}
\bibfield{author}{\bibinfo{person}{Jie Zhu}, \bibinfo{person}{Ying Shan},
  \bibinfo{person}{JC Mao}, \bibinfo{person}{Dong Yu}, \bibinfo{person}{Holakou
  Rahmanian}, {and} \bibinfo{person}{Yi Zhang}.}
  \bibinfo{year}{2017}\natexlab{}.
\newblock \showarticletitle{Deep embedding forest: Forest-based serving with
  deep embedding features}. In \bibinfo{booktitle}{\emph{Proceedings of the
  23rd ACM SIGKDD International Conference on Knowledge Discovery and Data
  Mining}}. ACM, \bibinfo{pages}{1703--1711}.
\newblock


\end{thebibliography}
